# Pump frequency resonances for light-induced incipient superconductivity in YBa$_2$Cu$_3$O$_{6.5}$

B. Liu[1], M. Först[1], M. Fechner[1], D. Nicoletti[1], J. Porras[3], T. Loew[3], B. Keimer[3], A. Cavalleri[1,2]

[1]*Max Planck Institute for the Structure and Dynamics of Matter, Hamburg, Germany.*
[2]*Department of Physics, University of Oxford, Clarendon Laboratory, Oxford, UK*
[3]*Max Planck Institute for Solid State Research, Stuttgart, Germany.*

**Optical excitation in the cuprates has been shown to induce transient superconducting correlations above the thermodynamic transition temperature, T$_C$, as evidenced by the terahertz frequency optical properties in the non-equilibrium state. In YBa$_2$Cu$_3$O$_{6+x}$ this phenomenon has so far been associated with the nonlinear excitation of certain lattice modes and the creation of new crystal structures. In other compounds, like La$_{2-x}$Ba$_x$CuO$_4$, similar effects were reported also for excitation at near infrared frequencies, and were interpreted as a signature of the melting of competing orders. However, to date it has not been possible to systematically tune the pump frequency widely in any one compound, to comprehensively compare the frequency dependent photo-susceptibility for this phenomenon. Here, we make use of a newly developed optical parametric amplifier, which generates widely tunable high intensity femtosecond pulses, to excite YBa$_2$Cu$_3$O$_{6.5}$ throughout the entire optical spectrum (3 – 750 THz). In the far-infrared region (3 – 25 THz), signatures of non-equilibrium superconductivity are induced only for excitation of the 16.4 THz and 19.2 THz vibrational modes that drive *c*-axis apical oxygen atomic positions. For higher driving frequencies (25 – 750 THz), a second resonance is observed around the charge transfer band edge at ~350 THz. These observations highlight the importance of coupling to the electronic structure of the CuO$_2$ planes, either mediated by a phonon or by charge transfer.**

The equilibrium superconducting state of high-$T_C$ cuprates manifests itself in a number of characteristic features in the terahertz-frequency optical response. In Figure 1, we report selected optical properties measured in $YBa_2Cu_3O_{6.5}$ above and below the superconducting transition temperature, $T_C$.

As the temperature is lowered from 100 K (black curve, $T \gg T_C \simeq 52$ K) to 10 K (red curve, $T \ll T_C$), the real part of the c-axis optical conductivity, $\sigma_1(\omega)$, evolves from that of a semiconductor with thermally-activated carriers to a gapped spectrum (see Fig. 1b.1). Simultaneously, a zero-frequency δ-peak emerges, indicative of dissipation-less DC transport. This peak is not seen directly in $\sigma_1(\omega)$, but is reflected in a $\sim 1/\omega$ divergence in the imaginary conductivity, $\sigma_2(\omega)$ (Fig. 1b.2). Correspondingly, a sharp edge in the optical reflectivity of the superconducting state develops at the Josephson Plasma Resonance (JPR) $\omega \simeq \omega_{JPR} \simeq 30$ cm$^{-1}$ [1,2,3] (Fig. 1b.3).

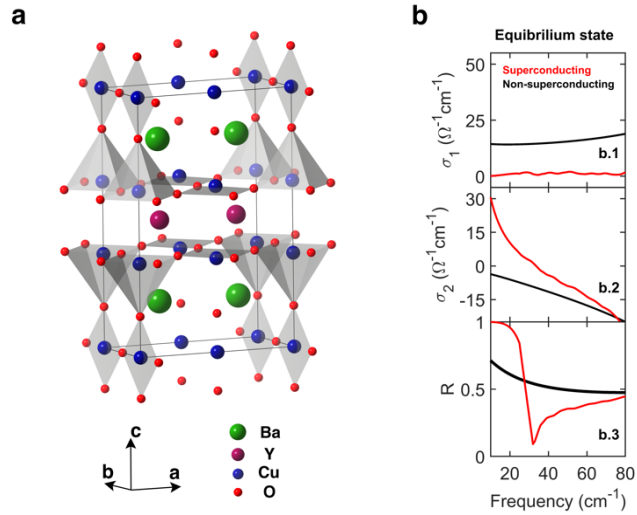

**FIG. 1. Crystal structure and equilibrium THz optical properties of $YBa_2Cu_3O_{6.5}$.** (a) Crystallographic unit cell of orthorhombic $YBa_2Cu_3O_{6.5}$. Bilayers of conducting $CuO_2$ planes, stacked along the c direction, form Josephson junctions in the superconducting state. (b) c-axis THz-frequency optical properties in the equilibrium superconducting ($T < T_C$, red curves) and normal ($T > T_C$, black curves) state. The real, $\sigma_1(\omega)$, and imaginary part, $\sigma_2(\omega)$, of the optical conductivity are displayed along with the normal-incidence reflectivity, $R(\omega)$ (same data as those reported in Refs. [4,5,6]).

A number of recent pump-probe experiments have shown that these same optical signatures (red curves in Fig. 1b) can be recreated transiently in $YBa_2Cu_3O_{6+x}$ for base temperatures $T \gg T_C$ by optical excitation made resonant with the 20-THz lattice vibrations that modulate the position of the apical oxygen atoms along the *c* axis [4,5,6]. Measurements of the transient atomic structure with femtoscond x-ray diffraction revealed an average structural deformation in $YBa_2Cu_3O_{6.5}$ [7], associated with nonlinear lattice dynamics [8,9]. It was reasoned that such a transient structure may favor higher temperature superconductivity [7].

However, the response of $YBa_2Cu_3O_{6+x}$ or that of any other material has never been systematically checked for excitation of different lattice modes, as no optical device existed that could generate high intensity pulses with sufficient spectral selectivity and tunability throughout the terahertz spectrum.

Furthermore, recent work in single-layer cuprates of the type $La_{2-x}Ba_xCuO_4$ [10,11,12,13,14] has evidenced transient optical properties similar to those observed for mid-infrared driving in $YBa_2Cu_3O_{6+x}$, for excitation in the near-infrared and visible part of the spectrum, a phenomenon that has been assigned to melting of a competing charge order [15,16] (as already discussed for $La_{1.675}Eu_{0.2}Sr_{0.125}CuO_4$ [17]).

Here we report a comprehensive study of the response of $YBa_2Cu_3O_{6.5}$ (the same compound investigated in Refs. 4,5,6,7) to excitation at all frequencies throughout the terahertz electromagnetic spectrum (3 – 24 THz; 100 – 800 cm$^{-1}$), as well as in the near-infrared and visible range (up to 750 THz). To this end, we make use of a newly developed nonlinear optical device [18] based on difference frequency mixing of chirped near infrared pulses [19] in organic crystals (see Supplemental Material [20]).

In a first set of experiments we studied the response to excitation of all phonons in the far infrared (3 – 24 THz; 100 – 800 cm$^{-1}$). The relevant pump frequency range is displayed in Fig. 2b, which reports the equilibrium broadband *c*-axis optical conductivity, $\sigma_1(\omega)$, at $T = 100$ K [2,3].

Here, we show data taken under the same conditions to those reported in Refs. 4,5,6, to be used as a reference point for all the experiments that follow. The excitation pulses were centered at 19.2 THz (640 cm$^{-1}$), which correspond to apical oxygen distortions at the oxygen deficient chains. These experiments were performed with the same broadband ~4-THz-wide pulses used in Refs. 4,5,6, and then, for comparison, with the newly available ~ 1-THz spectral bandwidth (narrowband). These pump pulses, polarized along the *c* axis of a YBa$_2$Cu$_3$O$_{6.5}$ single crystal [20], were focused onto the sample at a fluence of ~8 mJ/cm$^2$. Note that in the narrowband experiments the pulses were four times longer than in the broadband experiments (~600 fs vs. ~150 fs). As these two measurements used the same pump fluence, the peak electric fields were of 3 and 6 MV/cm for narrowband (long pulse) and broadband (short pulse) excitation, respectively.

As already discussed in Refs. 4,5,6, the transient *c*-axis optical properties were interrogated between ~15 and 80 cm$^{-1}$ by reflecting a second terahertz probe pulse generated by optical rectification in a nonlinear crystal. The electric field of these pulses, after reflection from the sample surface, was electro-optically sampled (Fig. 2b, grey spectrum) for different pump-probe time delays (see Supplemental Material [20] for further details).

Figures 2c-2d report the photoinduced changes in the complex optical conductivity at $T = 100\ K \gg T_C$, as a function of frequency and pump-probe time delay (color plots). Both excitation schemes induced qualitatively similar optical properties, with a significant increase in the imaginary conductivity, $\sigma_2(\omega)$, which became positive and exhibited a superconducting-like ~ $1/\omega$ divergence for $\omega \to 0$.

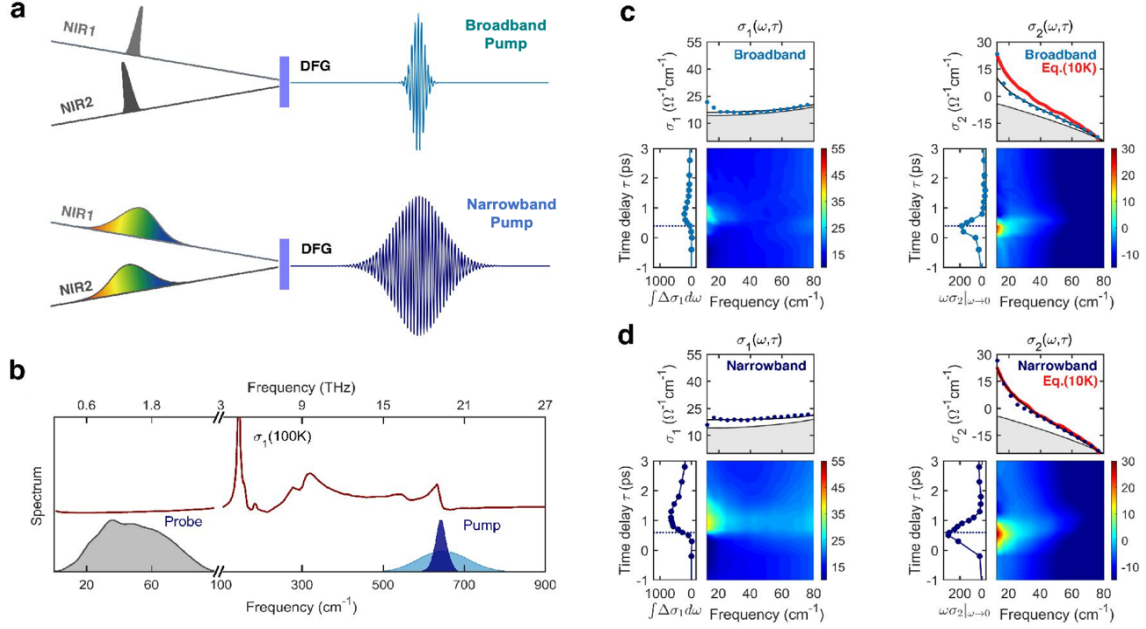

**FIG. 2. Transient *c*-axis optical conductivity induced by apical oxygen excitation above T$_C$.** (a) Schematics of the generation processes for broadband and narrowband excitation pulses. The signal outputs (NIR1 and NIR2) of two parallel optical parametric amplifiers were directly sent to a nonlinear crystal to generate broadband pump pulses (relative bandwidth Δν/ν ~ 20%) in a difference frequency generation (DFG) process. For narrowband generation (Δν/ν < 10%), the two signal outputs were linearly chirped before the DFG process (see Supplemental Material [20] for further details). (b) Equilibrium *c*-axis optical conductivity, $\sigma_1(\omega)$, of YB$_2$Cu$_3$O$_{6.5}$ at 100 K ($T > T_C$, red solid line), along with the frequency spectra of THz probe (grey), broadband pump (light blue), and 19.2 THz narrowband pump (dark blue) pulses. (c) Color plots: Frequency- and time-delay-dependent complex optical conductivity measured after broadband excitation at T = 100 K. Upper panels: corresponding $\sigma_1(\omega)$ and $\sigma_2(\omega)$ line cuts displayed at equilibrium (grey lines) and at the peak of the coherent response ($\tau \simeq 0.5$ ps time delay, blue circles). Black lines are fits to the transient spectra with a model describing the response of a Josephson plasma (see Supplemental Material [20] for details on the fitting procedure). For comparison, we also report the equilibrium $\sigma_2(\omega)$ measured in the superconducting state at T = 10 K (red line). Side panels: Frequency-integrated dissipative ($\int \Delta\sigma_1(\omega)d\omega$) and coherent ($\omega\sigma_2(\omega)|_{\omega\to 0}$) responses, as a function of pump-probe time delay. The delay corresponding to the spectra reported in the upper panels ($\tau \simeq 0.5$ ps) is indicated by a dashed line. (d) Same quantities as in (c), measured for narrowband excitation at 19.2 THz.

This is particularly evident in the frequency spectra measured at the peak of the signal, displayed above the color plots in Figs. 2c-2d (right panels). In these line cuts, the transient $\sigma_2(\omega)$ measured at $\tau \simeq 0.5$ ps after excitation (blue dots) is superimposed with the equilibrium $\sigma_2(\omega)$ at $T < T_C$ (red line) for comparison.

Note that in both these experiments, for which the excitation fluence was a factor of two higher than that reported in Refs. 4, 5, and 6, the response was that of a homogeneous medium. The effective medium model, introduced to reproduce the partial changes in the optical properties in Refs. 4 and 5, was no longer necessary to explain the data. We interpret this observation by positing that the excitation range, possible with the new setup, may have reached a dynamical "percolation threshold". Strikingly, for narrowband excitation at $T = 100$ K we observed exactly the same $\sigma_2(\omega)$ spectrum measured in the equilibrium superconducting state ($T = 10$ K $\ll T_C = 52$ K) in the same sample.

The photo-induced dynamics at longer time delays ($\tau \gtrsim 1$ ps) evidenced decoherence and increased dissipation, as observed in the real part of the optical conductivity, $\sigma_1(\omega)$ (Figs. 2c-2d, left panels). On the left hand side of each panel, we plot two frequency-integrated quantitites as a function of time delay: $\omega\sigma_2(\omega)|_{\omega\to 0}$, which in an equilibrium superconductor is proportional to the superfluid density, and $\int \Delta\sigma_1(\omega)d\omega$, which is a reporter of dissipation and quasiparticle heating [21]. For both broadband and narrowband excitation, it is evident that the dissipative part of the optical response ($\int \Delta\sigma_1(\omega)d\omega$, left panels in Figs. 2c-2d), increases only at later time delays compared to the superconducting component, reaching a maximum after the peak in $\omega\sigma_2(\omega)|_{\omega\to 0}$, has relaxed (right panels in Figs. 2c-2d).

Figure 3 reports a more comprehensive set of experiments, obtained by tuning the pump pulse frequency widely throughout the far-infrared spectrum. Four selected results are displayed, corresponding to resonant narrowband excitation of four different phonon modes (see Supplemental Material [20] for additional data sets). The data reported in Fig. 2 ($\omega_{pump} = 19.2$ THz = 640 cm$^{-1}$) is shown alongside the results for excitations at $\omega_{pump} = 16.4$ THz = 547 cm$^{-1}$, $\omega_{pump} = 10.1$ THz =337 cm$^{-1}$, and $\omega_{pump} = 4.2$

THz = 140 cm$^{-1}$, all driven by maintaining constant ∼3 MV/cm peak electric field strength.

The atomic displacements of these vibrational modes are displayed in Fig. 3a. The 4.2 THz mode involves motions of the barium atoms and of the apical oxygens, while the 10.1 THz mode is associated with a planar Cu-O buckling motion. The 16.4 THz and the 19.2 THz modes, whose individual responses to photo-excitation could not be separated with the previously available broadband pump pulses [4,5,6], involve almost exclusively oscillations of the apical oxygen atoms on oxygen-rich and oxygen-deficient Cu-O chains, respectively.

The four pump-probe experiments reported in Fig. 3b-3c, where we display selected spectra taken at the peak of the coherent response (at time delay $\tau \simeq 0.5$ ps), show that driving at the two highest frequencies modes (16.4 THz and 19.2 THz) induces a superconducting-like response ($\sigma_2(\omega) \propto 1/\omega$), for which the transient complex conductivity is fitted by a model describing the optical response of a Josephson plasma. On the other hand, excitation of the two low-frequency modes (4.2 THz and 10.1 THz) causes only a moderate increase in dissipation and no superconducting component. This observation could be well reproduced, for all time delays, by a Drude-Lorentz model for normal conductors (see Supplemental Material [20] for details on the fitting procedure).

The same experiments as those reported in Fig. 3 were systematically repeated for 42 pump frequencies throughout the far-infrared spectrum (3 – 24 THz), for which we report in Fig. 4 the results of the analysis of the transient optical properties at the time delay corresponding to the peak of the coherent response.

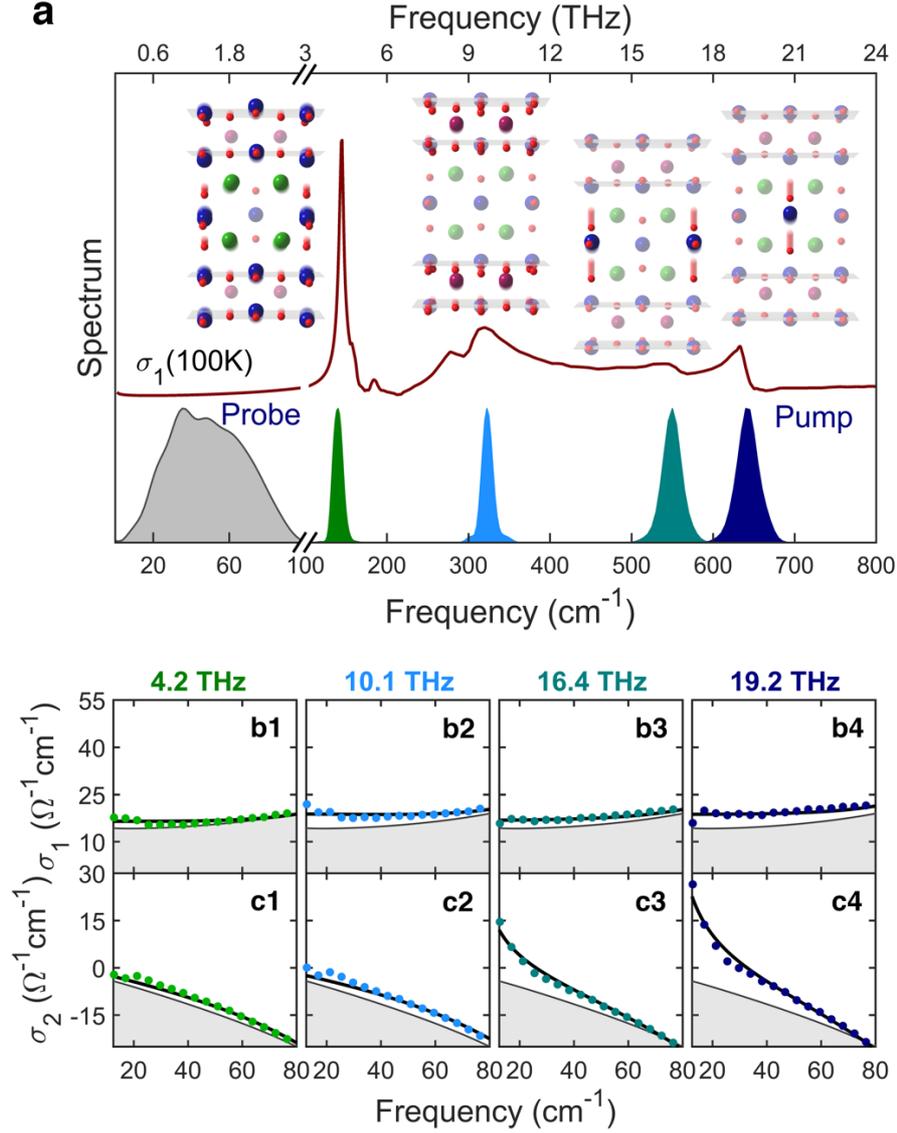

**FIG. 3. Transient *c*-axis optical response induced by mode-selective phonon excitations above T$_C$.** (a) Equilibrium *c*-axis optical conductivity, $\sigma_1(\omega)$, of YB$_2$Cu$_3$O$_{6.5}$ at 100 K ($T > T_C$, red solid line), along with the frequency spectra of THz probe (grey) and of different narrowband pump pulses, tuned to be resonant with four different optical phonons at 4.2 THz, 10.1 THz, 16.4 THz, and 19.2 THz. The insets sketch the atomic motions related to each of these modes, with CuO$_2$ layers highlighted in grey and transparency applied to quasi-stationary atoms (see Fig. 1 for the atom labeling). (b,c) Complex optical conductivity, $\sigma_1(\omega) + i\sigma_2(\omega)$, measured before (grey lines) and at $\tau \simeq 0.5$ ps time delay (colored circles) after resonant stimulation of the phonon modes shown in (a) with ~8 mJ/cm$^2$ pump fluence (corresponding to ~3 MV/cm peak electric field). The black solid lines are fits to the transient spectra performed with either a simple Drude-Lorentz model for normal conductors (b1, b2, c1, c2) or a model describing the response of a Josephson plasma (b3, b4, c3, c4) (see Supplemental Material [20] for more details on the fitting procedure).

In the top panel, we show the total, spectrally integrated probe signal, that is the modulus of the complex optical conductivity. In the middle plot, we display only the dissipative component of the signal ($\int \Delta\sigma_1(\omega)d\omega$), and in the lower plot only the superconducting contribution, $\omega\sigma_2(\omega)|_{\omega\to 0}$ (see Supplemental Material [20] for extended data sets).

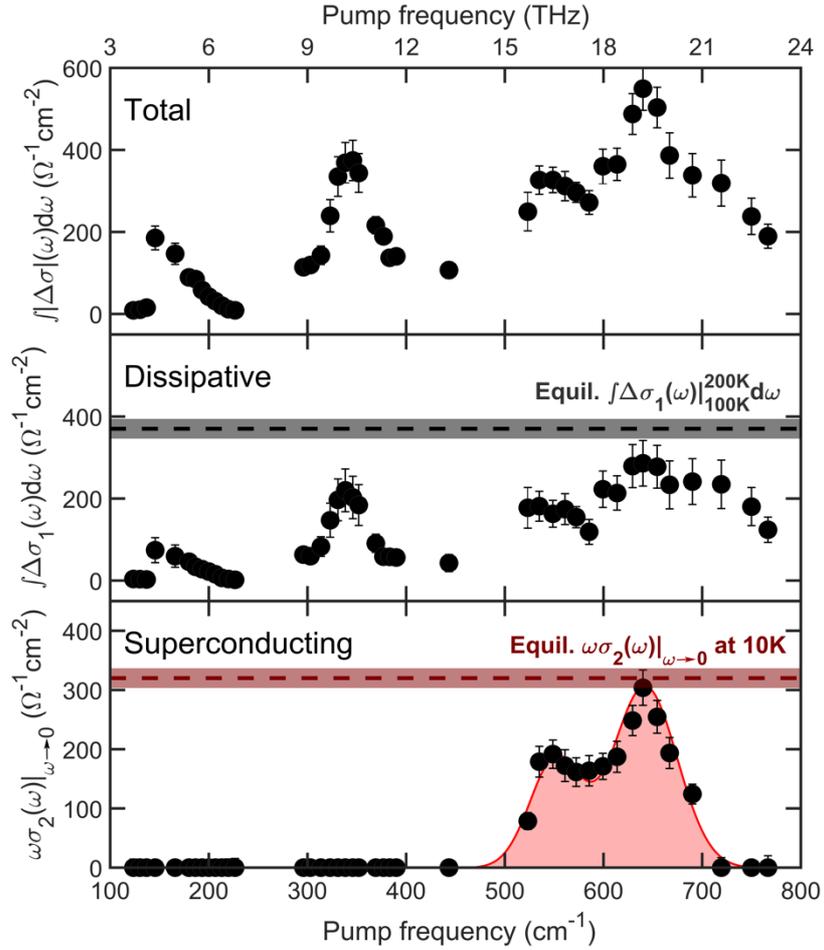

**FIG. 4. Evolution of the photo-induced response at T = 100 K as a function of excitation frequency.** Distinct quantities, extracted from the transient optical conductivity at $\tau \simeq 0.5$ ps time delay, are displayed as a function narrowband excitation frequency: (a) The total, frequency integrated change in optical conductivity $\int |\Delta\sigma(\omega)|d\omega$, (b) the frequency integrated dissipative response, $\int \Delta\sigma_1(\omega)d\omega$, and (c) the superconducting response represented by the low-frequency limit $\omega\sigma_2(\omega)|_{\omega\to 0}$. The red shaded region represents the frequency range around the apical oxygen phonon modes, where a transient superconducting-like response could be identified. Horizontal dashed lines indicate the thermally-induced increase in $\int \sigma_1(\omega)d\omega$ when heating the sample from 100 K to 200 K (black) and the equilibrium superfluid density, $\omega\sigma_2(\omega)|_{\omega\to 0}$, measured at T = 10 K (red). All data were taken with a pump fluence of ~8 mJ/cm² (corresponding to ~3 MV/cm peak electric field).

For comparison, we have also included horizontal dashed lines indicating the thermally-induced increase in $\int \sigma_1(\omega)d\omega$ when heating the sample from 100 K to 200 K (middle panel) and the equilibrium superfluid density, $\omega\sigma_2(\omega)|_{\omega\to 0}$, measured at T = 10 K (lower panel).

For excitation at the 19.2 THz discussed above and at 16.4 THz, the non-equilibrium state includes a dissipative $\int \Delta\sigma_1(\omega)d\omega$ response (analogous to that observed upon heating) that coexists with a superconducting-like imaginary conductivity ($\omega\sigma_2(\omega)|_{\omega\to 0}$) identical to that measured in the same material in the equilibrium superconducting state.

On the other hand, when the pump frequency is tuned below 15 THz (500 cm$^{-1}$), no superconducting-like component is observed, whereas the dissipative response is approximately the same as that observed for $\omega_{pump} > 15$ THz.

The different nature of the dissipative and the superconducting-like signal is underscored by the data reported in Figure 5. Here, we show the dependence of $\int \Delta\sigma_1(\omega)d\omega$ and $\omega\sigma_2(\omega)|_{\omega\to 0}$ on base temperature ($T > T_C$), for two different pump frequencies.

The dissipative term (Fig. 5a) is temperature-independent and persists all the way up to 325 K, which suggest that its origin may be related to heating of quasiparticles. On the other hand, the superconducting-like response (Fig. 5b), which is observed only for high-frequency pumping (blue circles), displays a strong reduction with increasing temperature, almost disappearing for T > 300 K. This is consistent with the observation reported in Refs. 5 and 6.

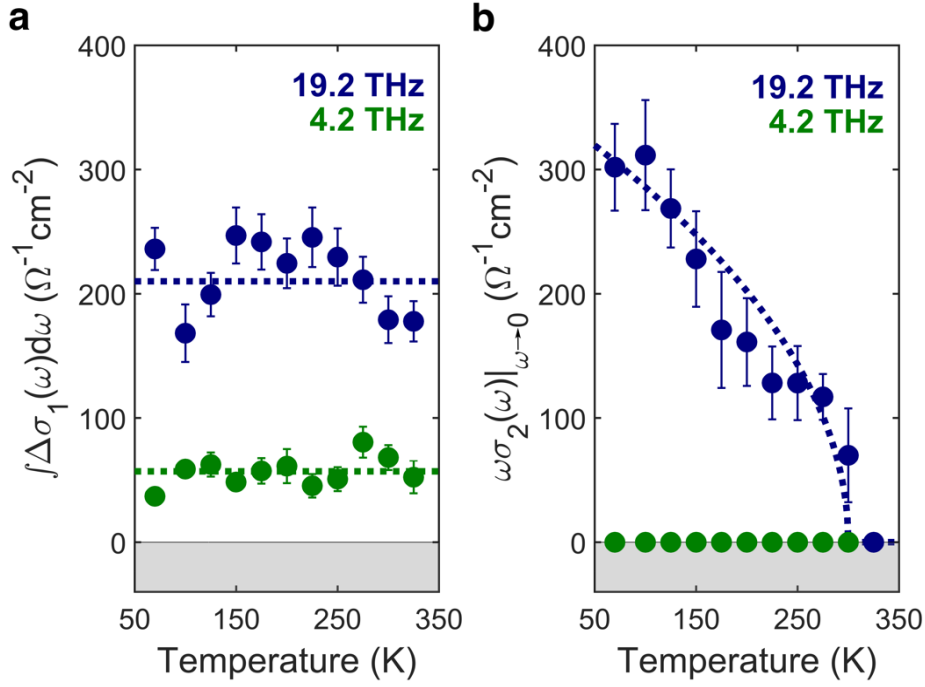

**FIG. 5. Temperature dependence of the dissipative and superconducting-like responses.** Temperature dependent (a) dissipative response, $\int \Delta\sigma_1(\omega)d\omega$, and (b) superconducting-like response, $\omega\sigma_2(\omega)|_{\omega \to 0}$, measured for resonant narrowband excitation of the lowest-frequency (4.2 THz) and highest-frequency (19.2 THz) phonon modes. The dashed lines are guides to the eye. As in Fig. 4, all data were taken with a pump fluence of ~8 mJ/cm² (corresponding to ~3 MV/cm peak electric field).

Additional data sets taken in the superconducting state at $T < T_C$ are reported in the Supplemental Material [20]. There, we show that the pre-existing, equilibrium superfluid density can be transiently enhanced (in agreement with Refs. 4 and 5) only by photoexcitation at 19.2 THz, provided that a sufficiently high pump fluence is employed. For weaker driving fields [22], or when the pump is detuned to lower frequencies, only a transient depletion of the superconducting condensate is observed. In a second set of experiments, we studied the response of the material to excitation at higher frequencies, above the phonon resonances and up to the region where electronic bands are found.

In Fig. 6a-6b we report similar measurements as those shown in Fig. 3 and display the transient complex conductivity at the peak of the light-induced response for three representative excitation frequencies: 29 THz ($\lambda_{pump} = 10.4$ µm), 214 THz ($\lambda_{pump} = 1.4$ µm), and 375 THz ($\lambda_{pump} = 800$ nm).

Similar to what reported in Fig. 4 for frequencies immediately above 19 THz, the response at 29 THz ($\lambda_{pump} = 10.4$ µm) displays no superconducting-like component. However, for excitation frequencies of 214 THz ($\lambda_{pump} = 1.4$ µm) and 375 THz ($\lambda_{pump} = 800$ nm) a transient $\sigma_2(\omega) \propto 1/\omega$ is observed again, resembling that induced by driving resonant with the apical oxygen phonons.

Note that for these higher frequencies, the reconstruction procedure is less reliable than for the experiments reported in Figs. 2-5, as the penetration depth mismatch of pump and probe pulses becomes larger. Indeed, at all times, the raw response (optically induced change of the reflected THz probe electric field) is largest at the two phonon resonances displayed in Figs. 3-4 [20]. The error bars for high excitation frequencies (see Fig. 6c) reflect this uncertainty. Nevertheless, error analysis indicates that the appearance of a divergent imaginary conductivity is robust.

A complete pump frequency dependence for the superconducting component $\omega\sigma_2(\omega)|_{\omega\to 0}$ is displayed in Figure 6c (see Supplemental Material [20] for extended data sets). A negligible response of the imaginary conductivity is found for all driving frequencies between the apical oxygen phonon (~ 21 THz) and ~ 42 THz, whereas for higher pump frequencies a second resonance emerges, with a strength that follows the increase in pump absorption on the charge transfer resonance (visualized in this figure by the equilibrium optical conductivity, $\sigma_1^{equil}(\omega_{pump})$).

Figure 6d displays the same data after normalization against the oscillator strength of the material at each pump frequency. The normalized quantity $\omega_{probe}\sigma_2(\omega_{probe})|_{\omega\to 0} / \sigma_1^{equil}(\omega_{pump})$ is the frequency dependent "photo-susceptibility" for the transient state and therefore to be taken as an alternative way of visualizing the efficiency of the effect.

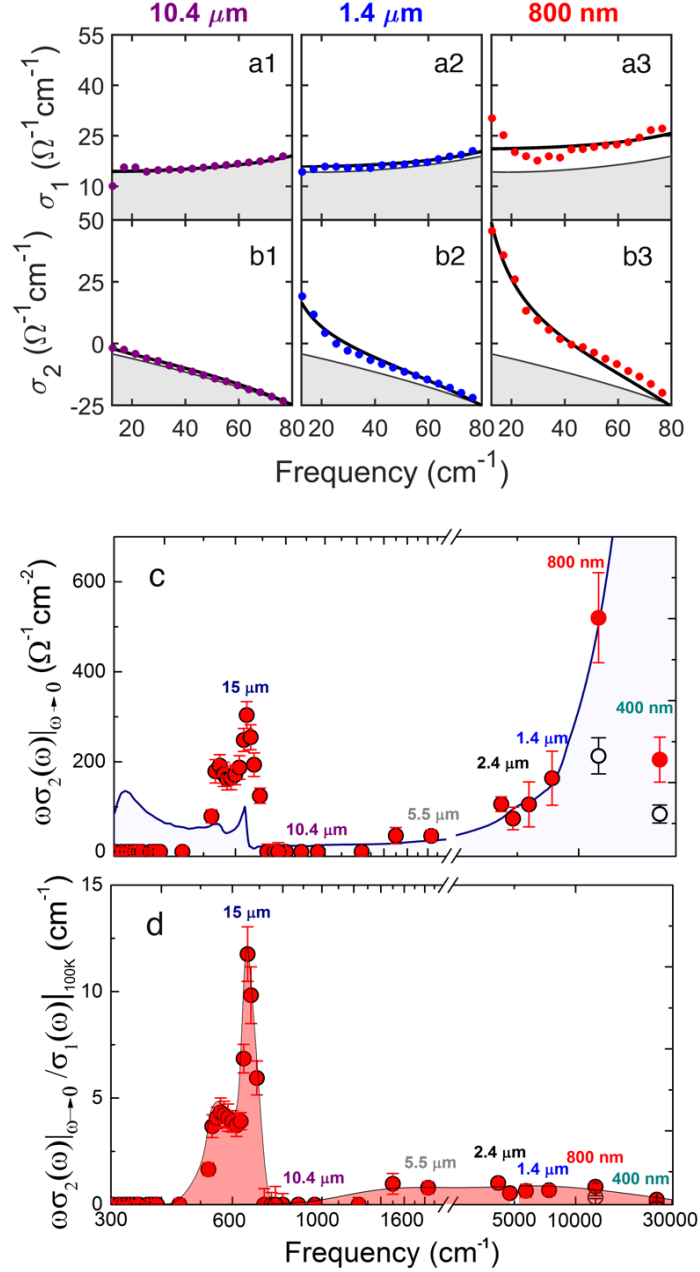

**FIG. 6. Transient *c*-axis optical response induced by high-energy charge excitation at T = 100 K.** (a,b) Complex optical conductivity, $\sigma_1(\omega) + i\sigma_2(\omega)$, measured at equilibrium (grey) and at $\tau \simeq 0.5$ ps time delay (colored circles) after excitation at three different pump wavelengths with ~8 mJ/cm² pump fluence. The black solid lines are fits to the transient spectra performed with either a simple Drude-Lorentz model for normal conductors (a1, b1) or a model describing the response of a Josephson plasma (a2, b2, a3, b3) (see Supplemental Material [20] for more details on the fitting procedure). (c) Pump frequency dependence of the superconducting response, represented by the low-frequency limit $\omega\sigma_2(\omega)|_{\omega\to 0}$ (red circles). The equilibrium optical conductivity, $\sigma_1^{equil}(\omega_{pump})$, is shown as a blue line. (d) Normalized response, $\omega\sigma_2(\omega)|_{\omega\to 0}/\sigma_1^{equil}(\omega_{pump})$ (red circles). The grey line is a multi-Lorentzian fit. All data for $\omega_{pump} < 300$ THz were taken with constant pump fluence (~8 mJ/cm²) and peak electric field (~3 MV/cm) at a fixed pulse duration of ~600 fs. At $\lambda_{pump} = 800$ nm and $\lambda_{pump} = 400$ nm, for which the pump pulses were ~100 fs long, we report data points taken both at ~8 mJ/cm² (~7 MV/cm, red circles) and ~1.5 mJ/cm² (~3 MV/cm, empty circles), keeping either the excitation fluence or peak electric field at the same values.

In the aggregate, the body of work reported above prompts the following considerations. Firstly, from the data in Figs. 3-5 it is clear that some form of mode-specific lattice excitation must underpin optically-induced superconductivity in the terahertz frequency range. In previous studies [7,9], it was conjectured that the lowest order nonlinear lattice anharmonicity, of the type $Q_{IR}^2 Q_R$ (where $Q_{IR}$ and $Q_R$ are the normal coordinates of the directly driven infrared-active mode and of any anharmonically coupled Raman mode, respectively [8]), may explain the observed phenomenology. Indeed, this lattice term leads to a transient, average structural deformation that may be beneficial to superconductivity.

Hence, in addition to the nonlinear phononic mechanism, which is validated by x-ray experiments and certainly present, other phenomena are likely to come into play. As documented in the Supplemental Material [20], when considering the calculated average lattice deformations induced by an anharmonic $Q_{IR}^2 Q_R$ coupling for driving each of the *IR* phonons of Figure 3, one does not find a defining feature for the two modes at 16.4 THz and 19.2 THz.

Note also that the two high-frequency vibrations drive large amplitude motions of the apical oxygen atoms, which are then expected to couple directly to the in-plane electronic and magnetic structure. This coupling is supposed to be much weaker for the other modes at lower frequency [23,24]. We also note that the frequency of the two apical oxygen phonons matches approximately the sum of the inter- and intra-bilayer Josephson plasma frequencies in YBa$_2$Cu$_3$O$_{6.5}$ ( $\omega_{JPR,1} \simeq 1 - 2$ THz and $\omega_{JPR,2} \simeq 14$ THz, respectively). Hence, a mechanism in which driven lattice excitations couple directly to the in-plane electronic structure may become resonantly enhanced at these frequencies [25,26,27,28,29].

The second resonance found at the charge transfer band appears to reinforce the notion that changes in the electronic properties of the planes is key in the observed phenomenon [30,31]. At the doping level studied here, charge-order melting is not expected to play an important role, as is the case of $YBa_2Cu_3O_{6.6}$ [32] and for $La_{1.875}Ba_{0.125}CuO_4$ [33]. At these frequencies, for the excitation polarized along the *c* axis, we expect a rearrangement of the electronic structure with some qualitative analogy to the direct action of the apical oxygen modes. It is therefore possible that a similar mechanism to that responsible for the resonances at 19.2 and 16.4 THz is at play. We have noted that for the two apical oxygen modes parametric excitation of interlayer fluctuations would be resonant with the sum frequency of the intra and inter bilayer modes [29]. Here, a similar parametric coupling may be at play, without the frequency resonance and hence less efficient.

Clearly, further studies that make use of the new pump device available here are needed, with special attention to measurements of time dependent lattice dynamics [7] and inelastic excitations [34,35,36]. More generally, the tuneable, spectrally-selective nonlinear pump source, applied for the first time in the present study, is expected to strongly impact the investigation of non-equilbrium phenomena in solids.

**REFERENCES (Main Text)**

[1] D. N. Basov and T. Timusk, "Electrodynamics of high-$T_c$ superconductors," *Rev. Mod. Phys.* **77**, 721 (2005).

[2] C. C. Homes, T. Timusk, D. A. Bonn, R. Liang, and W. N. Hardy. "Optical properties along the c-axis of $YBa_2Cu_3O_{6+x}$ for $x = 0.50 \rightarrow 0.95$ Evolution of the pseudogap" *Physica C* **254**, 265-280 (1995).

[3] C. C. Homes, T. Timusk, D. A. Bonn, R. Liang, and W.N. Hardy, "Optical phonons polarized along the *c*-axis of $YBa_2Cu_3O_{6+x}$, for x $\rightarrow$ 0.5 to 0.95," *Can. J. Phys.* **73**, 663 (1995).

[4] W. Hu, S. Kaiser, D. Nicoletti, C. R. Hunt, I. Gierz, M. C. Hoffmann, M. Le Tacon, T. Loew, B. Keimer, and A. Cavalleri, "Optically enhanced coherent transport in $YBa_2Cu_3O_{6.5}$ by ultrafast redistribution of interlayer coupling," *Nat. Mater*. **13**, 705 (2014).

[5] S. Kaiser, C. R. Hunt, D. Nicoletti, W. Hu, I. Gierz, H. Y. Liu, M. Le Tacon, T. Loew, D. Haug, B. Keimer, and A. Cavalleri, "Optically induced coherent transport far above $T_c$ in underdoped $YBa_2Cu_3O_{6+\delta}$," *Phys. Rev. B* **89**, 184516 (2014).

[6] C. R. Hunt, D. Nicoletti, S. Kaiser, D. Pröpper, T. Loew, J. Porras, B. Keimer, and A. Cavalleri, "Dynamical decoherence of the light induced interlayer coupling in $YBa_2Cu_3O_{6+\delta}$," *Phys. Rev. B* **94**, 224303 (2016).

[7] R. Mankowsky, A. Subedi, M. Först, S.O. Mariager, M. Chollet, H. Lemke, J. Robinson, J. Glownia, M. Minitti, A. Frano, M. Fechner, N. A. Spaldin, T. Loew, B. Keimer, A. Georges, and A. Cavalleri, "Nonlinear lattice dynamics as a basis for enhanced superconductivity in $YBa_2Cu_3O_{6.5}$", *Nature* **516**, 71 (2014).

[8] M. Först, C. Manzoni, S. Kaiser, Y. Tomioka, Y. Tokura, R. Merlin, and A. Cavalleri, "Nonlinear phononics as an ultrafast route to lattice control", *Nat. Phys.* **7**, 854 (2011).

[9] R. Mankowsky, M. Först, T. Loew, J. Porras, B. Keimer, and A. Cavalleri, "Coherent modulation of the $YBa_2Cu_3O_{6+x}$ atomic structure by displacive stimulated ionic Raman scattering", *Phys. Rev. B* **91**, 094308 (2015).

[10] D. Nicoletti, E. Casandruc, Y. Laplace, V. Khanna, C. R. Hunt, S. Kaiser, S. S. Dhesi, G. D. Gu, J. P. Hill, and A. Cavalleri, "Optically-induced superconductivity in striped $La_{2-x}Ba_xCuO_4$ by polarization-selective excitation in the near infrared", *Phys. Rev. B* **90**, 100503(R) (2014).

[11] E. Casandruc, D. Nicoletti, S. Rajasekaran, Y. Laplace, V. Khanna, G. D. Gu, J. P. Hill, and A. Cavalleri, "Wavelength-dependent optical enhancement of superconducting interlayer coupling in $La_{1.885}Ba_{0.115}CuO_4$", *Phys. Rev. B* **91**, 174502 (2015).

[12] D. Nicoletti, D. Fu, O. Mehio, S. Moore, A. S. Disa, G. D. Gu, and A. Cavalleri, "Magnetic-Field Tuning of Light-Induced Superconductivity in Striped $La_{2-x}Ba_xCuO_4$", *Phys. Rev. Lett.* **121**, 267003 (2018).

[13] K. A. Cremin, J. Zhang, C. C. Homes, G. D. Gu, Z. Sun, M. M. Fogler, A. J. Millis, D. N. Basov, and R. D. Averitt, "Photo-enhanced metastable c-axis electrodynamics in stripe ordered cuprate $La_{1.885}Ba_{0.115}CuO_4$", *Proc. Natl. Acad. Sci. U.S.A*. **116**, 19875-19879 (2019).

[14] S. J. Zhang, Z. X. Wang, L. Y. Shi, T. Lin, M. Y. Zhang, G. D. Gu, T. Dong, and N. L. Wang, "Light-induced new collective modes in $La_{1.905}Ba_{0.095}CuO_4$ superconductor", *Phys. Rev. B* **98**, 020506(R) (2018).

[15] M. Först, R.I. Tobey, H. Bromberger, , S.B. Wilkins, V. Khanna, A.D. Caviglia, Y.D. Chuang, W.S. Lee, W.F. Schlotter, J.J. Turner, M.P. Minnitti, O. Krupin, X.J, Xu, J.S. Wen, G.D. Gu, S. S. Dhesi, A. Cavalleri, and J. P. Hill "Melting Charge Stripes in Vibrationally Driven

La$_{1.875}$Ba$_{0.125}$CuO$_4$: assessing the respective roles of Electronic and Lattice order in frustrated Supeconductors", *Phys. Rev. Lett.* **112**, 157002 (2014).

[16] V. Khanna, R. Mankowsky, M. Petrich, H. Bromberger, S. A. Cavill, E. Möhr-Vorobeva, D. Nicoletti, Y. Laplace, G. D. Gu, J. P. Hill, M. Först, A. Cavalleri and S. S. Dhesi, "Restoring interlayer Josephson coupling in La$_{1.885}$Ba$_{0.115}$CuO$_4$ by charge transfer melting of stripe order", *Phys. Rev. B* **93**, 224522 (2016).

[17] D. Fausti, R.I. Toby, N. Dean. S. Kaiser, A. Dienst, M. Hoffmann, S. Pyon, T. Takayam, H. Takagi, and A. Cavalleri " Light induced superconductivity in a striped ordered cuprate", *Science* **331**, 189 (2011).

[18] B. Liu, H. Bromberger, A. Cartella, T. Gebert, M. Först, and A. Cavalleri, "Generation of narrowband high-intensity, carrier-envelope phase stable pulses tunable between 4 and 18 THz", *Opt. Lett.* **42**, 129-131 (2017).

[19] A. Cartella, T. F. Nova, A. Oriana, G. Cerullo, M. Först, C. Manzoni, and A. Cavalleri, "Narrowband carrier-envelope phase stable mid-infrared pulses at wavelengths beyond 10 μm by chirped-pulse difference frequency generation", *Opt. Lett.* **42**, 663-666 (2017).

[20] See Supplemental Material for details on the experimental setup, data acquisition and evaluation, equilibrium optical properties, uncertainties in the determination of the transient optical properties, fitting models, extended data sets, pump electric field dependence, transient response below T$_C$, pump induced heating, and the ab-initio calculations of the structural dynamics.

[21] M. Dressel and G. Grüner, Electrodynamics of Solids, Cambridge University Press, Cambridge (2002)

[22] S. J. Zhang, Z. X. Wang, H. Xiang, X. Yao, Q. M. Liu, L. Y. Shi, T. Lin, T. Dong, D. Wu, and N. L. Wang, "Photo-induced nonequilibrium response in underdoped YBa$_2$Cu$_3$O$_{6+x}$ probed by time-resolved terahertz spectroscopy", *arXiv:1904.10381* (2019).

[23] M. Mori, G. Khalliullin, T. Tohyama, and S. Maekawa, "Origin of the spatial variation of the pairing gap in Bi-based high temperature superconductors", *Phys. Rev. Lett.* **101**, 24720023 (2008).

[24] Y.Y . Peng, G. Dellea, M. Conni, A. Amorese, D. Di Castro, G.M. DeLuca, K. Kummer, M. Salluzzo, X. Sun, X.J. Zhou, G. Balestino, M. LeTacon, B. Keimer, L. Braicovic, N.B. Brookes, and G. Ghiringhelli "Influence of apical oxygen on the extent of the in-plane exchange interaction in cuprate superconductors" *Nat. Phys.* **13**, 1201 (2017).

[25] S. J. Denny, S. R. Clark, Y. Laplace, A. Cavalleri, and D. Jaksch, "Proposed Parametric Cooling of Bilayer Cuprate Superconductors by Terahertz Excitation", *Phys. Rev. Lett.* **114**, 137001 (2015).

[26] J. Okamoto, A. Cavalleri, and L. Mathey, "Theory of enhanced interlayer tunneling in optically driven high T$_c$ superconductors", *Phys. Rev. Lett.* **117**, 227001 (2016).

[27] M. Knap, M. Babadi, G. Refael, I. Martin, and E. Demler, "Dynamical Cooper pairing in nonequilibrium electron-phonon systems", *Phys. Rev. B* **94**, 214504 (2016).

[28] M. Babadi, M. Knap, I. Martin, G. Refael, and E. Demler, "Theory of parametrically amplified electron-phonon superconductivity", *Phys. Rev. B* **96**, 014512 (2017).

[29] A. von Hoegen, M. Fechner, M. Först, J. Porras, B. Keimer, M. Michael, E. Demler, and A. Cavalleri, "Probing coherent charge fluctuations in YBa$_2$Cu$_3$O$_{6+x}$ at wavevectors outside the light cone", *arXiv:1911.08284* (2019).

[30] G. Yu, C. H. Lee, A. J. Heeger, N. Herron, E. M. McCarron, L. Cong, G. C. Spalding, C. A. Nordman, and A. M. Goldman, "Phase separation of photogenerated carriers and photoinduced superconductivity in high-$T_C$ materials", *Phys. Rev. B* **45**, 4964 (1992).

[31] G. Nieva, E. Osquiguil, J. Guimpel, M. Maenhoudt, B. Wuyts, Y. Bruynseraede, M. B. Maple, and I. K. Schuller, "Photoinduced enhancement of superconductivity", *Appl. Phys. Lett.* **60**, 2159 (1992).

[32] M. Först, A. Frano, S. Kaiser, R. Mankowsky, C. R. Hunt, J. J. Turner, G. L. Dakovski, M. P. Minitti, J. Robinson, T. Loew, M. Le Tacon, B. Keimer, J. P. Hill, A. Cavalleri, and S. S. Dhesi, "Femtosecond x rays link melting of charge-density wave correlations and light-enhanced coherent transport in $YBa_2Cu_3O_{6.6}$", *Phys. Rev. B* **90**, 184514 (2014).

[33] M. Först, R.I. Tobey, H. Bromberger, , S.B. Wilkins, V. Khanna, A.D. Caviglia, Y.D. Chuang, W.S. Lee, W.F. Schlotter, J.J. Turner, M.P. Minnitti, O. Krupin, X.J, Xu, J.S. Wen, G.D. Gu, S. S. Dhesi, A. Cavalleri and J. P. Hill "Melting Charge Stripes in Vabrationally Driven $La_{1.875}Ba_{0.125}CuO_4$: assessing the respective roles of Electronic and Lattice order in frustrated Supeconductors", *Phys. Rev. Lett.* **112**, 157002 (2014).

[34] M. Buzzi, M. Först, R. Mankowsky, and A. Cavalleri, "Probing dynamics in quantum materials with femtosecond X-rays", *Nat. Rev. Mater.* **3**, 299-311 (2018).

[35] L. J. P. Ament, M. van Veenendaal, T. P. Devereaux, J. P. Hill, and J. van den Brink, "Resonant inelastic x-ray scattering studies of elementary excitations", *Rev. Mod. Phys.* **83**, 705 (2011).

[36] H.-H. Kim, S.M. Souliou, M. E. Barber, E. Lefrançois, M.Minola, M. Tortora, R. Heid, N. Nandi, R. A. Borzi, G. Garbarino, A. Bosak, J. Porras, T. Loew, M. König, P. J.W. Moll, A. P. Mackenzie, B. Keimer, C. W. Hicks, and M. Le Tacon, "Uniaxial pressure control of competing orders in a high-temperature superconductor", *Science* **362**, 1040 (2018).

# Pump Frequency Resonances for Light-Induced Incipient Superconductivity in $YBa_2Cu_3O_{6.5}$

B. Liu[1], M. Först[1], M. Fechner[1], D. Nicoletti[1], J. Porras[3], T. Loew[3], B. Keimer[3], A. Cavalleri[1,2]

[1]Max Planck Institute for the Structure and Dynamics of Matter, 22761 Hamburg, Germany.
[2]Department of Physics, University of Oxford, Clarendon Laboratory, Oxford OX1 3PU, UK
[3]Max Planck Institute for Solid State Research, 70569 Stuttgart, Germany.

# Supplemental Material

**S1. Experimental setup**

**S2. Data acquisition and evaluation**

**S3. Equilibrium optical properties**

**S4. Uncertainties in the determination of the transient optical properties**

**S5. Fitting models**

**S6. Extended data sets**

**S7. Pump electric field dependence**

**S8. Transient response below $T_C$**

**S9. Pump-induced heating**

**S10. Ab-initio calculations of the structural dynamics**

**References**

# S1. Experimental setup

The optical setup used to photoexcite $YBa_2Cu_3O_{6.5}$ with tunable narrowband pump pulses and to probe its low-frequency THz response is sketched in Figure S1. It was fed by a 1-kHz repetition rate Ti:sapphire regenerative amplifier, delivering 7-mJ, 80-fs pulses at 800 nm wavelength.

We used 90% of the pulse energy to generate the excitation pulses. To this end, we pumped two parallel home-built optical parametric amplifiers (OPA), which were seeded with the same white light to produce carrier-envelope-phase (CEP) stable THz pump pulses in a subsequent difference frequency generation (DFG) process [1,2]. The two OPA signal outputs of ~120 fs duration were directly sent to a 1-mm thick GaSe crystal to generate the broadband pump pulses at 19.2 THz with relative bandwidth $\Delta\nu/\nu \sim 20\%$. For the generation of narrowband pulses with relative bandwidth $\Delta\nu/\nu < 10\%$, the two signal outputs were linearly chirped to a pulse duration of ~600 fs before the DFG process in either a 0.5 mm thick DSTMS organic crystal [3] or in the same 1-mm GaSe crystal used for the broadband pulses. The pulse energies achieved were always at the few-µJ level. The maximum peak electric fields of the focused beams at the sample surface were 2.9 and 2.7 MV/cm for 4.2 and 10 THz, respectively, and about 3.1 MV/cm for both 16.4 and 19.2 THz pulses. They could be attenuated by inserting a pair of free-standing wire grid polarizers.

Excitation pulses with frequencies above the phonon resonances and up to the near-infrared and visible light regime were either generated by difference-frequency mixing the two OPA outputs in the GaSe crystal, or were taken directly as the output of one of the OPAs, as the fundamental 800 nm pulses from the Ti:sapphire amplifier, or as their second harmonic generated via type-I phase matching in a BBO crystal.

Single-cycle THz probe pulses with spectral components between 10 and 80 cm$^{-1}$ were generated via optical rectification of 500-µJ pulses at the fundamental 800 nm wavelength in a 500-µm thick ZnTe crystal. The THz electric field of the probe pulses reflected from the sample surface was detected via electro-optic sampling in a 500-µm thick optically contacted ZnTe crystal.

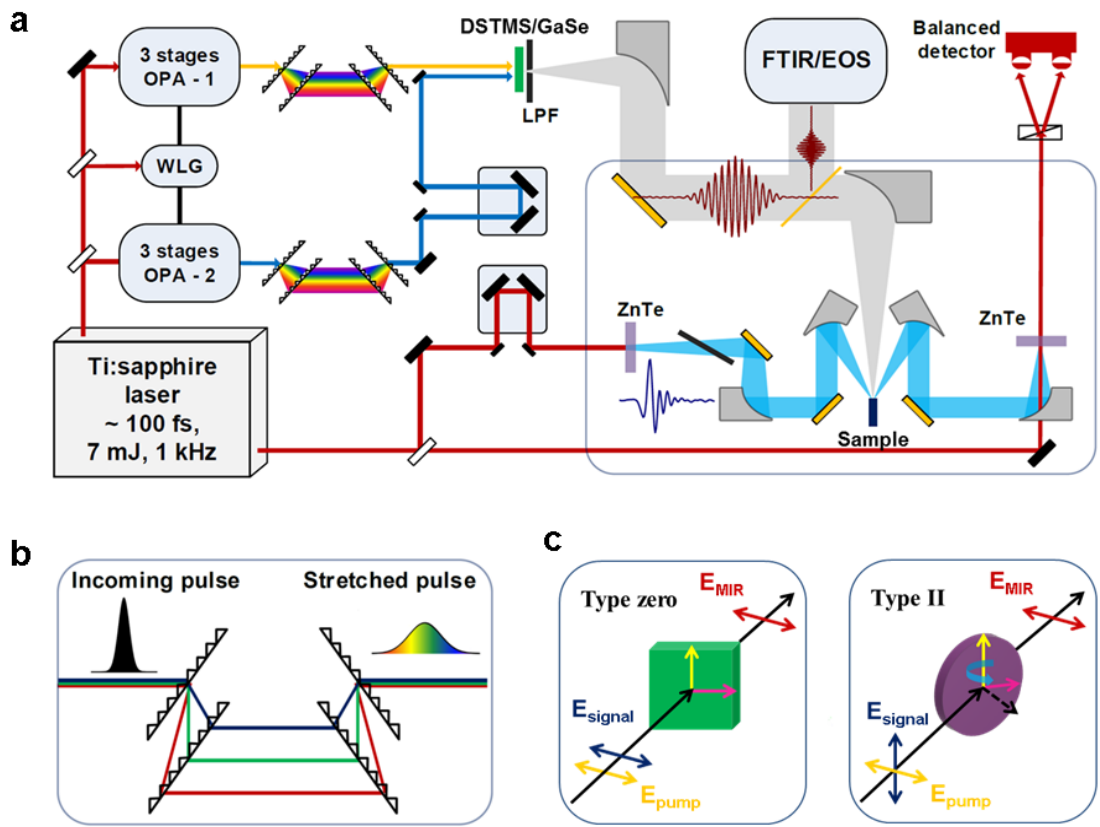

**Figure S1. a.** Narrowband pump / THz probe setup. OPA, optical parametric amplifier. WLG, white light generation. LPF, long-pass filter. Two near-infrared signal outputs from two parallel OPAs were linearly chirped and sent to a nonlinear crystal (DSTMS/GaSe) for difference frequency generation (DFG). The generated pump pulses were separated from the near-infrared inputs and characterized by either Fourier-transform infrared spectroscopy (FTIR) or electro-optic sampling (EOS). These pump pulses were set to be polarized along the $c$ axis of YBa$_2$Cu$_3$O$_{6.5}$, with their frequency tuned to be resonant with different phonon modes. The transient optical properties were measured by detecting the THz electric field generated from a ZnTe crystal via optical rectification and reflected from the sample surface, as a function of pump-probe time delay. **b.** Pulse stretcher consisting of four transmission gratings. **c.** Phase-matching condition and DFG process in DSTMS (left) and GaSe (right) crystals.

The experiments were performed on the *ac* surface of a YBa$_2$Cu$_3$O$_{6.5}$ single crystal. Pump and probe beams were always polarized along the crystallographic *c* axis. The crystal dimension along this axis (~300 μm) was smaller than the spot size of the pump beam (~400 μm), thus ensuring to probe a homogeneously excited region, independent on the probe spot dimension. A sharp superconducting transition at T$_C$ = 52 K was determined by dc magnetization measurements [4].

## S2. Data acquisition and evaluation

Time-domain THz spectroscopy was used to characterize the transient response of YBa$_2$Cu$_3$O$_{6.5}$ induced by optical driving. The spectral response at each time delay after excitation was obtained by keeping fixed the delay $\tau$ between the pump pulse and the electro-optic sampling gate pulse, and scanning the single-cycle THz probe pulse with internal delay *t* across.

The stationary probe electric field $E_R(t)$ and the differential electric field $\Delta E_R(t,\tau)$ reflected from the sample were recorded simultaneously by feeding the electro-optic sampling signal into two lock-in amplifiers and mechanically chopping the pump and probe beams at different frequencies of ~357 and 500 Hz, respectively. The differential signal $\Delta E_R(t,\tau)$ was sampled at the ~143 Hz difference frequency of the inner and outer wheels of the same optical chopper. This approach minimized the cross-talk between the two detected signals whilst reducing the noise level of the measurements.

The electric field $E_R(t)$ and the differential field $\Delta E_R(t,\tau)$ were independently Fourier transformed to obtain the complex-valued, frequency-dependent $\tilde{E}_R(\omega)$ and $\Delta\tilde{E}_R(\omega,\tau)$. The photo-excited complex reflection coefficient $\tilde{r}(\omega,\tau)$ was determined by [5,6]

$$\frac{\Delta \tilde{E}_R(\omega,\tau)}{\tilde{E}_R(\omega)} = \frac{\tilde{r}(\omega,\tau) - \tilde{r}_0(\omega)}{\tilde{r}_0(\omega)},$$

where $\tilde{r}_0(\omega)$ is the stationary reflection coefficient known from the equilibrium optical response (see Section S3).

The penetration depth of the excitation pulses was typically smaller than that of the low-frequency THz probe pulses, implying that we were not probing a homogeneously excited sample volume. This mismatch was considered in the data analysis. At each frequency, the penetration depths were calculated by $d(\omega) = \frac{c}{2\omega \cdot \text{Im}[\tilde{n}_0(\omega)]}$ (here $\tilde{n}_0(\omega)$ is the stationary complex refractive index), yielding values of ~5–15 $\mu$m for the THz probe and in the range of ~0.1-5 $\mu$m for the pump pulses (see also Section S3).

An accurate estimate of the photo-induced optical response functions was then achieved by treating the sample as a layered system, in which only a thin layer below the sample surface is homogeneously excited while the bulk layer below remains unperturbed. The complex reflection coefficient is then expressed as

$$\tilde{r}(\omega,\tau) = \frac{\tilde{r}_A(\omega,\tau) + \tilde{r}_B(\omega)e^{2i\delta(\omega,\tau)}}{1 + \tilde{r}_A(\omega,\tau)\tilde{r}_B(\omega)e^{2i\delta(\omega,\tau)}}$$

Here, $\tilde{r}_A(\omega,\tau)$ and $\tilde{r}_B(\omega,\tau)$ are the reflection coefficients at the interfaces vacuum/photoexcited layer and photoexcited layer/unperturbed bulk, respectively, while $\delta = 2\pi d \tilde{n}(\omega,\tau)/\lambda_0$, with $\tilde{n}(\omega,\tau)$ being the complex refractive index of the photo-excited layer and $\lambda_0$ the probe wavelength.

Numerically solving this equation allowed us to retrieve $\tilde{n}(\omega,\tau)$ from the reflection coefficient $\tilde{r}(\omega,\tau)$ obtained in the experiment. The complex optical conductivity for the homogeneously excited volume of the material is expressed as $\tilde{\sigma}(\omega,\tau) = \frac{\omega}{4\pi i}[\tilde{n}(\omega,\tau)^2 - \varepsilon_\infty]$, with $\varepsilon_\infty = 4.5$ as a standard value for high-$T_C$ cuprates.

In Figure S2.1 & S2.2 we report, for specific data sets, all the individual steps involved in the determination of the transient optical response functions.

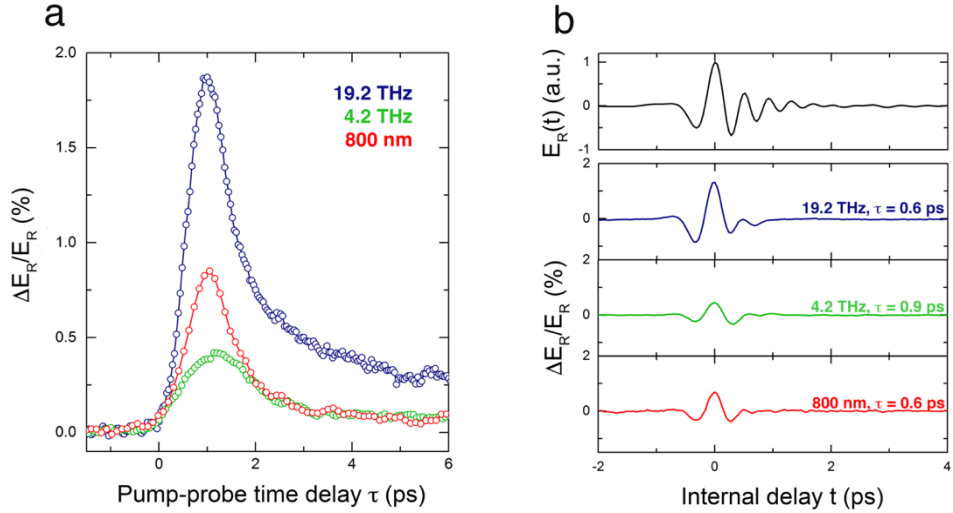

**Figure S2.1. a.** Raw pump-induced changes at the peak of the THz probe electric field (internal delay $t = 0$) as a function of pump-probe delay $\tau$, measured in YBa$_2$Cu$_3$O$_{6.5}$ at T = 100 K for excitation at $\omega_{pump}$ = 4.2 THz (green), $\omega_{pump}$ = 19.2 THz (blue), and $\lambda_{pump}$ = 800 nm (red). **b**. Top panel: Electro-optic sampling scan of the stationary THz probe electric field profile, $E_R(t)$, reflected from the sample. Lower panels: Transient scans of the pump-induced changes in $E_R(t)$, measured for the same pump frequencies as in panel (a), at a selected pump-probe time delay, $\tau$.

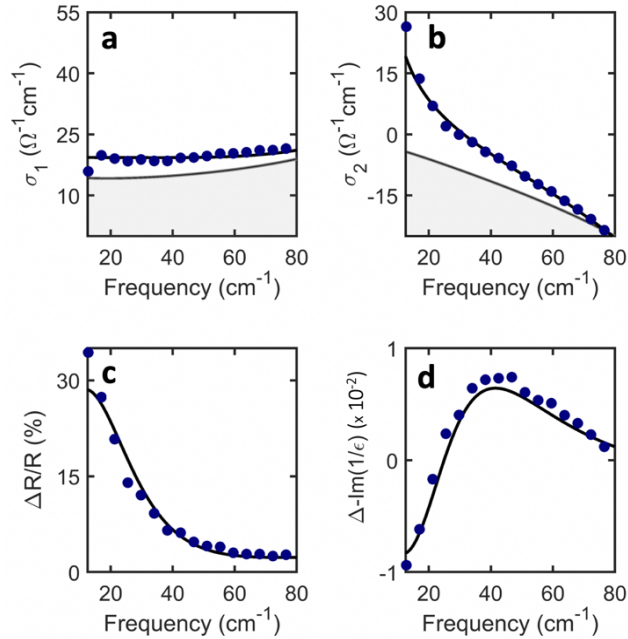

**Figure S2.2.** Real and imaginary part of the optical conductivity (**a, b**), relative changes in the reflectivity (**c**) and in the loss function (**d**), calculated for a homogeneously excited volume from the raw data of Fig. S2.1 ($\omega_{pump}$ = 19.2 THz, time delay $\tau$ = 0.6 ps), using the penetration depth mismatch model described in the text (blue circles). Black lines are fits to the transient spectra (see Section S5), while grey lines indicate the same quantities measured in equilibrium.

## S3. Equilibrium optical properties

The equilibrium optical properties of $YBa_2Cu_3O_{6.5}$ were determined following the same procedure described in Refs. 4 & 5.

The stationary reflected electric field, $E_R(t)$, was measured with the time-domain THz spectroscopy setup described in Sections S1 & S2, at different temperatures below and above $T_C$ = 52 K. This was then Fourier transformed to obtain the complex-valued, frequency dependent $\tilde{E}_R(\omega)$.

The equilibrium reflectivity in the superconducting state, $R(\omega, T < T_C)$, was determined as $R(\omega, T < T_C) = \frac{|\tilde{E}_R(\omega, T<T_C)|^2}{|\tilde{E}_R(\omega, T\gtrsim T_C)|^2} R(\omega, T \gtrsim T_C)$. Here, $R(\omega, T \gtrsim T_C)$ is the normal-state reflectivity measured with Fourier-transform infrared spectroscopy [7], which is weakly frequency and temperature dependent in the THz range. An example of this procedure is shown in Fig. S3.1.

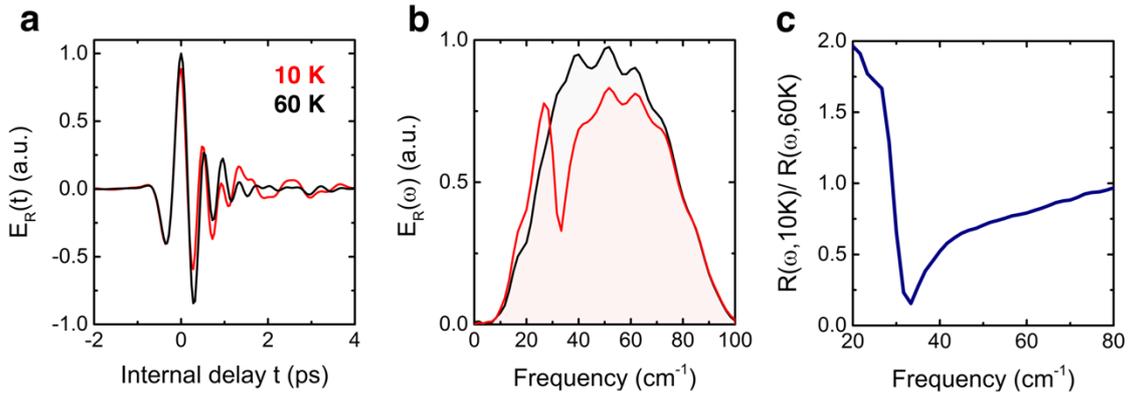

**Figure S3.1.** **a**. Stationary electric field profile measured after reflection from $YBa_2Cu_3O_{6.5}$ at $T < T_C$ (red) and $T = 60\ K \gtrsim T_C$ (black). **b**. Fourier transform amplitude of both quantities shown in panel a. **c**. Reflectivity ratio extracted from the spectra in panel (b), showing a Josephson Plasma Resonance at $\omega \simeq$ 35 cm[-1], in agreement with literature data [7].

These THz-frequency reflectivities were then fitted with the same models discussed in Section S5 (Josephson Plasma and Drude-Lorentz models for below-$T_C$ and above-$T_C$ data, respectively) and merged at $\omega \simeq 80$ cm$^{-1}$ with the broadband spectra from Refs. [7,8,9]. This allowed us to perform Kramers-Kronig transformations, thus retrieving full sets of equilibrium optical response functions (*i.e.*, complex optical conductivity $\tilde{\sigma}_0(\omega)$, complex dielectric function $\tilde{\varepsilon}_0(\omega)$, complex refractive index $\tilde{n}_0(\omega)$ ) for all temperatures investigated in our pump-probe experiment. In Figure S3.2 we report selected quantities obtained with this procedure.

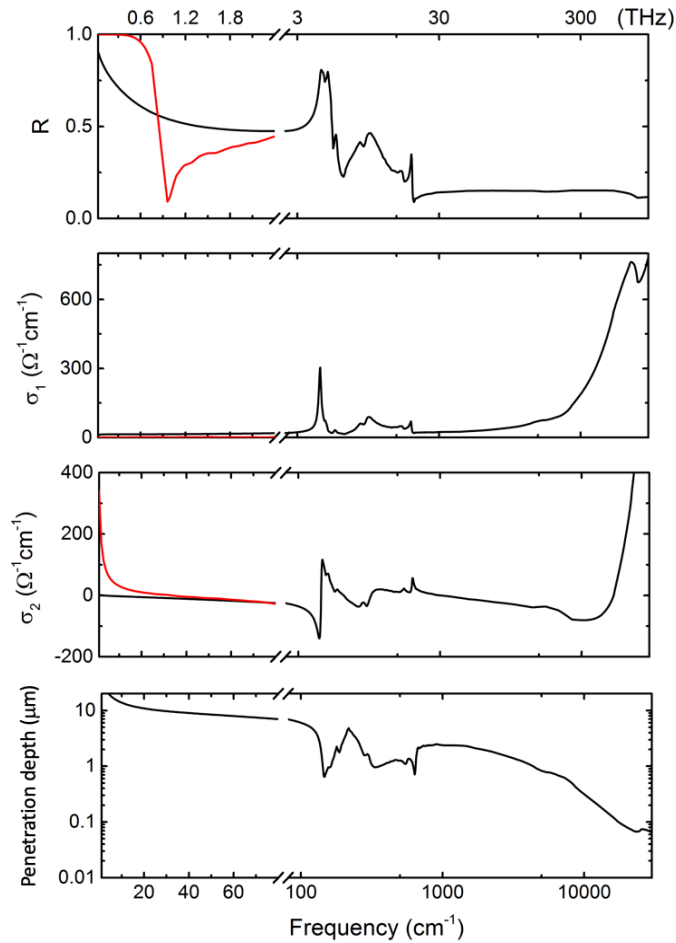

**Figure S3.2.** Equilibrium reflectivity, complex optical conductivity and penetration depth, $d(\omega) = \frac{c}{2\omega \cdot \mathrm{Im}[\tilde{n}_0(\omega)]}$, of YBa$_2$Cu$_3$O$_{6.5}$ at T = 100 K (black curves), extracted with the procedure described in the text. Red lines refer to below T$_C$ data (T = 10 K).

# S4. Uncertainties in the determination of the transient optical properties

We examine here different sources of uncertainty and their propagation in the data analysis. We choose to apply this examination to a representative data set taken at T = 100 K for narrowband excitation at 19.2 THz.

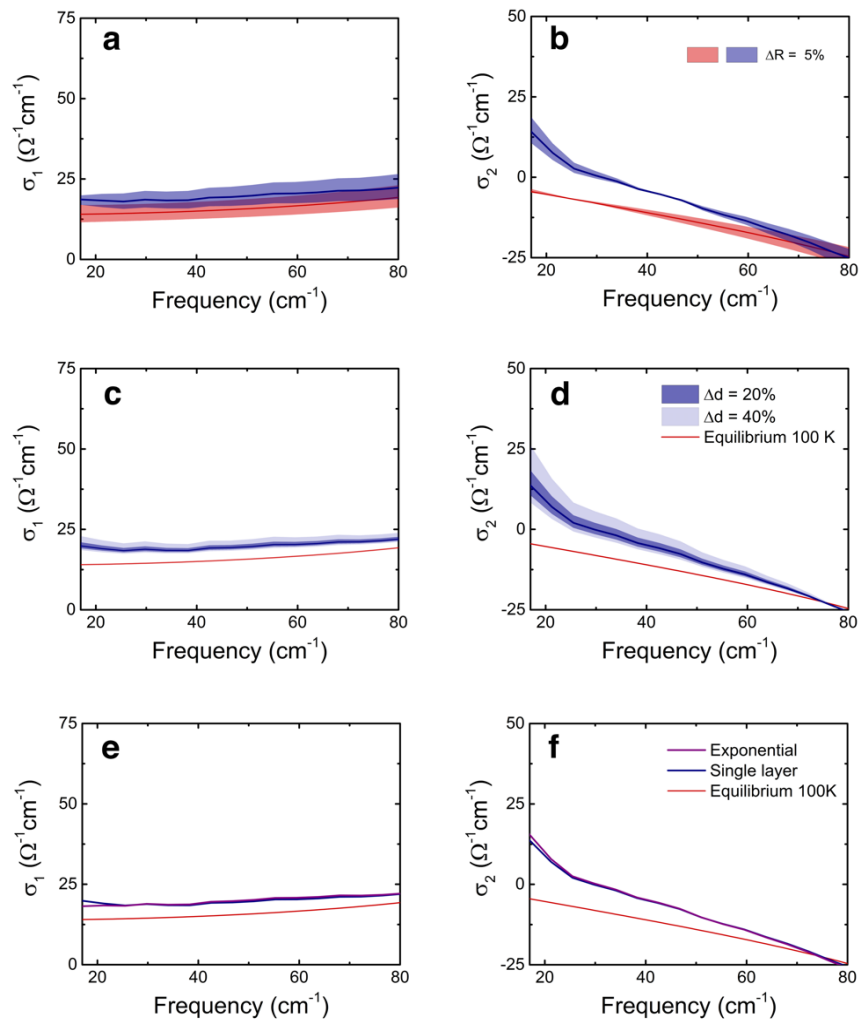

**Figure S4.** Effects of different sources of uncertainties in the determination of the transient optical conductivity of YBa$_2$Cu$_3$O$_{6.5}$ at T = 100 K and $\tau = 0.6$ ps time delays after narrowband excitation at 19.2 THz. In all panels, equilibrium values are shown in red, while transient spectra are in blue. Error bars, displayed as colored bands, have been propagated as follows: **a-b**, ±5% uncertainty in the equilibrium R(ω); **c-d**, ±20% and ±40% change in the pump penetration depth d= 0.7 $\mu$m. In **e-f** we analyze the effect of different functional forms for modelling the pump–probe penetration depth mismatch: a single-layer model, or a multi-layer model with exponential profile, both with the same pump penetration depth d= 0.7 $\mu$m.

The typical uncertainty in the absolute value of the measured equilibrium reflectivity is of the order of ±2-3%. Figure S4a-b shows as colored bands the propagated error bars in the equilibrium and transient optical properties for a ±5% uncertainty in R(ω). The response appears to be only marginally affected.

Another possible source of error resides in the value of the pump penetration depth used in the layered-system analysis to determine the transient response of a homogeneously excited volume. For 19.2 THz, this value was set to d= 0.7 $\mu$m. In Fig. S4c-d we show how a ±20% and a ±40% change in the pump penetration depth would affect the transient optical properties.

Finally, in Fig. S4e-f we show the effect of a different choice of the model for describing the pump probe penetration depth mismatch. The "single" photoexcited layer description, used for all data analyzed in the paper and described in Section S2, is compared with an exponential "multi-layer" decay model (used, for example, in Ref. [5]). Importantly, in all cases considered here, the impact of the different sources of error on the calculated optical conductivities is negligible.

## S5. Fitting models

At each pump probe time delay $\tau$, the transient *c*-axis optical response functions were fitted either with a model including the response of a Josephson plasma or with a simple Drude-Lorentz model for normal conductors [6]. A single set of fit parameters was always used to simultaneously reproduce the real and imaginary part of the optical conductivity, $\sigma_1(\omega)$ and $\sigma_2(\omega)$.

The phonon modes in the mid-infrared ( 150 cm$^{-1}$ ≲ $\omega$ ≲ 700 cm$^{-1}$) and the high-frequency electronic absorption ( $\omega$ ≳ 3000 cm$^{-1}$) were constructed by fitting the equilibrium spectra with Lorentz oscillators, for which the complex dielectric function is expressed as:

$$\tilde{\varepsilon}_{HF}(\omega) = \sum_i \frac{S_i^2}{(\Omega_i^2 - \omega^2) - i\omega\Gamma_i}.$$

Here, $\Omega_i$, $S_i$, and $\Gamma_i$ are the central frequency, strength, and damping coefficient of the *i*-th oscillator, respectively. In all fitting procedures to the transient data, these high-frequency terms were kept fixed.

The low-frequency Drude contribution to the complex dielectric function is expressed as

$$\tilde{\varepsilon}_D(\omega) = \varepsilon_\infty - \omega_P^2/(\omega^2 + i\Gamma\omega),$$

where $\omega_P$ and $\Gamma$ are the Drude plasma frequency and momentum relaxation rate, which were left as free fitting parameters, while $\varepsilon_\infty$ was kept fixed to 4.5.

The dielectric function of the Josephson plasma is expressed instead as

$$\tilde{\varepsilon}_J(\omega) = \varepsilon_\infty - \omega_J^2/\omega^2 + \tilde{\varepsilon}_N(\omega).$$

Here, the free fit parameters are the Josephson plasma frequency, $\omega_J$, and $\tilde{\varepsilon}_N(\omega)$, a weak "normal fluid" component [6,10] (overdamped Drude term), which was introduced to reproduce the positive offset in $\sigma_1(\omega)$.

As an example, Figure S5 reports the photo-induced optical conductivity for 19.2 THz narrowband excitation at different time delays τ = 0.6 ps and 2.4 ps and the corresponding fits. At early time delay (τ = 0.6 ps), the Josephson plasma model (black solid lines) is able to reproduce the experimental data. Notably, a Drude response alone (dashed lines) would not capture the measured low-frequency increase in $\sigma_2(\omega)$.

At later time delays (Fig. S5b), for which the system has already evolved into a dissipative response, the experimental data could instead be well reproduced by the simple Drude model for normal conductors.

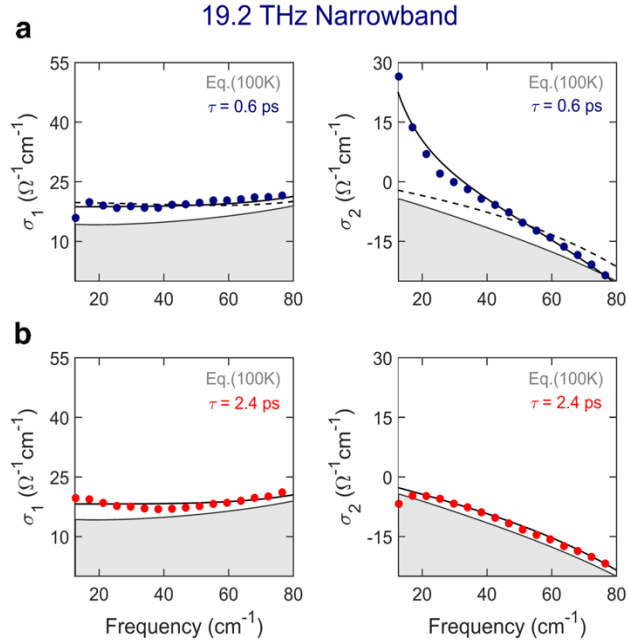

**Figure S5.** Fits to the transient complex optical conductivity measured after narrowband excitation at $\omega_{pump}$ = 19.2 THz (T = 100 K). **a.** Real and imaginary part of the optical conductivity measured at τ = 0.6 ps pump-probe time delay (blue dots). Grey curves with shading are the corresponding equilibrium spectra (shown also in panels b). The black solid line is a fit with the Josephson plasma model. The dashed black curve is instead a failed fit attempt using a simple Drude term for normal conductors. **b.** Real and imaginary conductivity measured at τ = 2.4 ps time delay (red dots). This dissipative response could be well reproduced by the Drude-Lorentz model (black solid lines).

## S6. Extended data sets

The full dynamical evolution of the photoinduced changes in the complex optical conductivity at $T = 100\,K \gg T_C$, which is reported in Fig. 2 of the main text for both broadband and narrowband excitation at 19.2 THz, is shown here also for narrowband excitation at 16.4 THz (Fig. S6.1), 4.2 THz, and 10.1 THz (Fig. S6.2).

For each data set, we display color plots of the real and imaginary part of the optical conductivity as a function of both frequency and pump-probe time delay, along with selected spectra measured at the peak of the response (in analogy with Fig. 2 in the main text). In addition, we plot two frequency-integrated quantities as a function of time delay: $\omega\sigma_2(\omega)|_{\omega\to 0}$, which in a superconductor is proportional to the superfluid density, and $\int \Delta\sigma_1(\omega)d\omega$, which is a reporter of dissipation and quasiparticle heating inside the gap.

As already extensively discussed in the main text, only driving at the two high-frequency modes at 16.4 THz and 19.2 THz induced a superconducting-like response ($\sigma_2(\omega) \propto 1/\omega$), for which the transient complex conductivity was fitted by a model describing the optical response of a Josephson plasma. On the other hand, excitation of the two low-frequency modes (at 4.2 THz and 10.1 THz) caused a moderate increase in dissipation and no superconducting component. This observation could be well reproduced, for all time delays, by a simple Drude-Lorentz model for normal conductors.

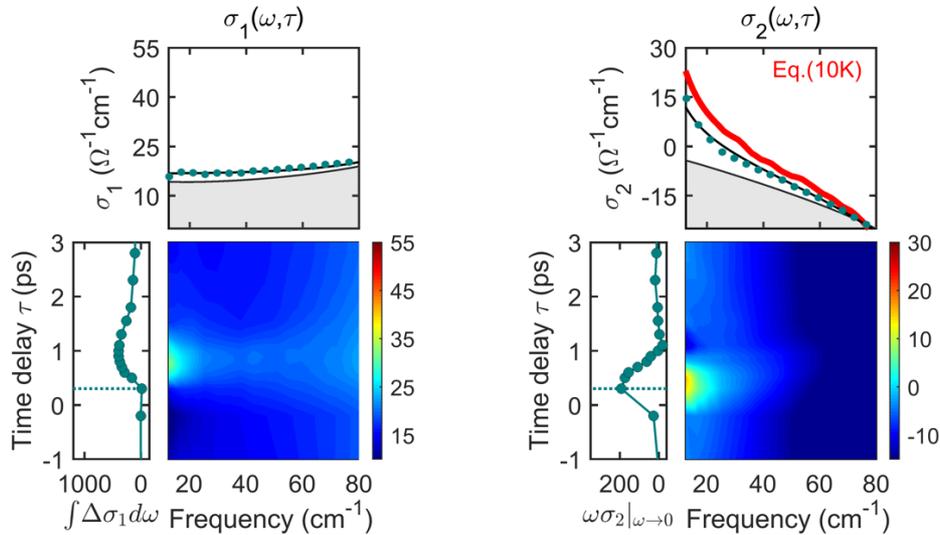

**Figure S6.1**. Frequency- and time-delay-dependent complex optical conductivity measured at T = 100 K for narrowband excitation at 16.4 THz (color plots). Upper panels: corresponding $\sigma_1(\omega)$ and $\sigma_2(\omega)$ line cuts displayed at equilibrium (grey lines) and at the time delay corresponding to the peak of the coherent response (teal circles). Black lines are fits to the transient spectra with a model describing the response of a Josephson plasma. For comparison, we also report the equilibrium $\sigma_2(\omega)$ measured in the superconducting state at T = 10 K (red line). Side panels: Frequency-integrated dissipative ($\int \Delta\sigma_1(\omega)d\omega$) and coherent ($\omega\sigma_2(\omega)|_{\omega\to 0}$) responses, as a function of pump-probe time delay. The delay corresponding to the spectra reported in the upper panels is indicated by a dashed line.

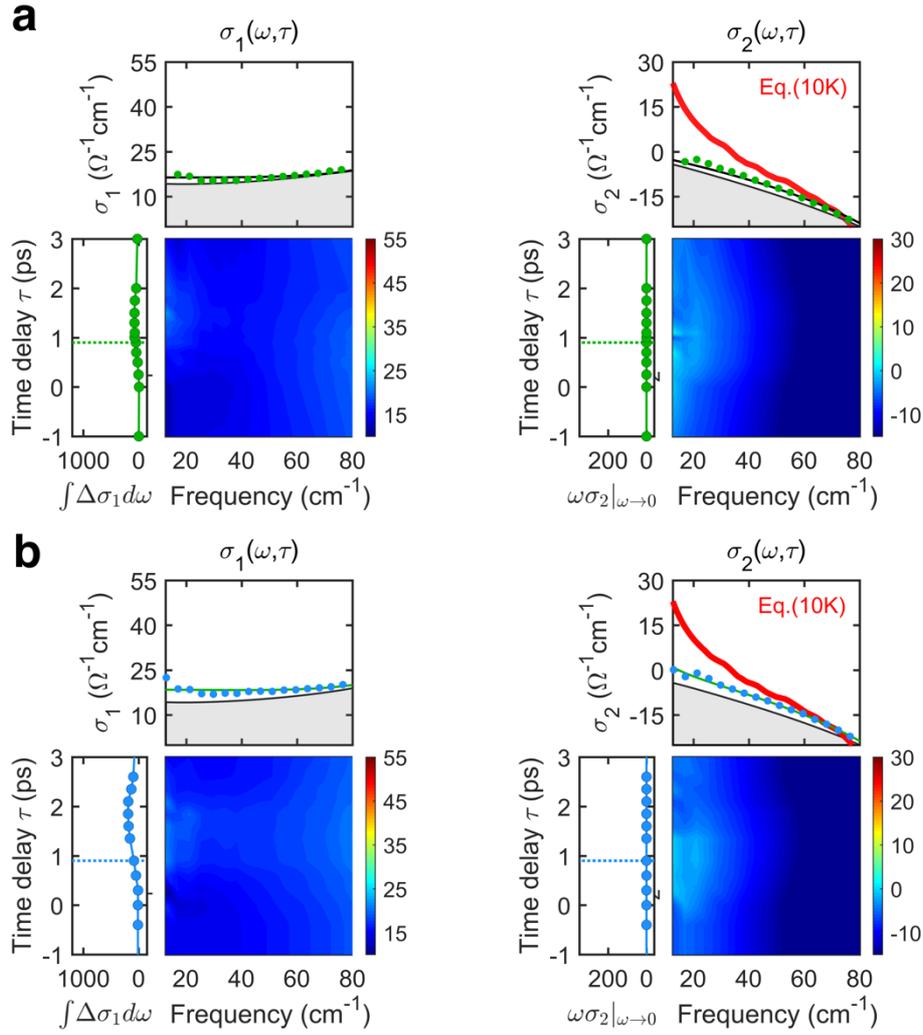

**Figure S6.2**. **a.** Frequency- and time-delay-dependent complex optical conductivity measured at T = 100 K for narrowband excitation at 4.2 THz (color plots). Upper panels: corresponding $\sigma_1(\omega)$ and $\sigma_2(\omega)$ line cuts displayed at equilibrium (grey lines) and at the time delay corresponding to the peak of the response (green circles). Black lines are fits to the transient spectra with a Drude-Lorentz model. For comparison, we also report the equilibrium $\sigma_2(\omega)$ measured in the superconducting state at T = 10 K (red line). Side panels: Frequency-integrated dissipative ($\int \Delta\sigma_1(\omega)d\omega$) and coherent ($\omega\sigma_2(\omega)|_{\omega\to 0}$) responses, as a function of pump-probe time delay. The delay corresponding to the spectra reported in the upper panels is indicated by a dashed line. **b.** Same quantities as in (a), measured for narrowband excitation at 10.1 THz.

In another set of experiments, we have also investigated the low-frequency THz response of YBa$_2$Cu$_3$O$_{6.5}$ for excitation above the phonon resonances, up to the near-infrared and visible spectral range.

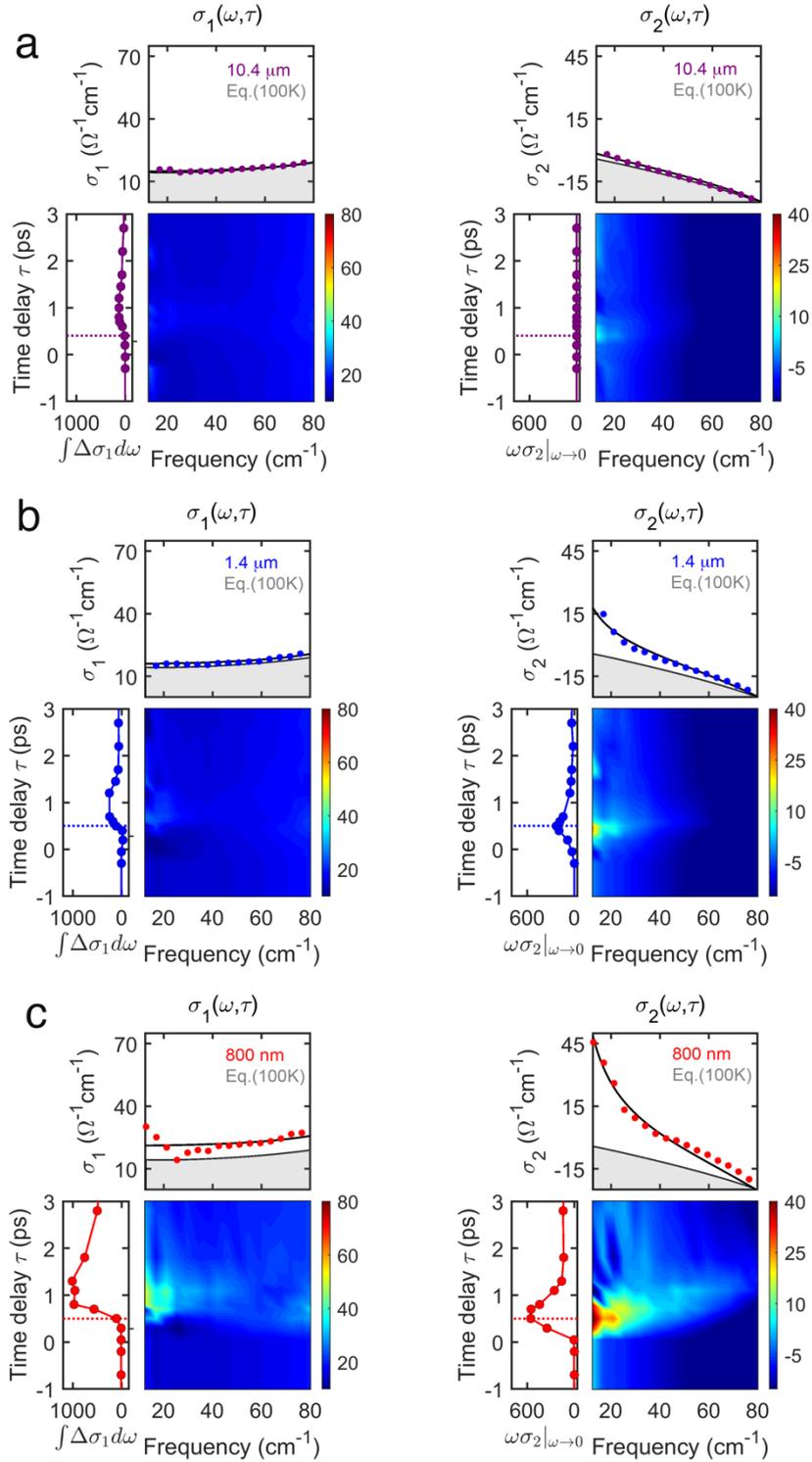

**Figure S6.3**. **a.** Frequency- and time-delay-dependent complext optical conductivity measured at T = 100 K for excitation at 10.4 μm (color plots). Upper panels: corresponding $\sigma_1(\omega)$ and $\sigma_2(\omega)$ line cuts displayed at equilibrium (grey lines) and at the time delay corresponding to the peak of the response (purple circles). Black lines are fits to the transient spectra with a Drude-Lorentz model. Side panels: Frequency-integrated dissipative ($\int \Delta\sigma_1(\omega)d\omega$) and coherent ($\omega\sigma_2(\omega)|_{\omega\to 0}$) responses, as a function of pump-probe time delay. The delay corresponding to the spectra reported in the upper panels is indicated by a dashed line. **b-c.** Same quantities as in (a)**,** measured for excitation at 1.4 μm and 800 nm.

The transient complex conductivity measured at the peak of the light-induced response for three selected excitation frequencies, 29 THz ($\lambda_{pump}$ = 10.4 μm), 214 THz ($\lambda_{pump}$ = 1.4 μm), and 375 THz ($\lambda_{pump}$ = 800 nm), is reported in Fig. 6 of the main text.

The full dynamical evolution of the photoinduced changes in the complex optical conductivity for these three excitation wavelengths and a fixed excitation fluence of ~8 mJ/cm² is shown here in Figs. S6.3. For each data set, we display the same quantities as in Figs. S6.1 and S6.2.

Remarkably, whilst the response to 10.4 μm driving is marginal, excitation at the two highest frequencies induces a transient optical conductivity compatible with dissipation-less transport ($\sigma_2(\omega) \propto 1/\omega$), strongly resembling that induced by driving resonant with the apical oxygen phonons (see also the discussion in the main text).

## S7. Pump electric field dependence

In Figure S7 we report the dependence of the coherent component of the pump-induced response, $\omega\sigma_2(\omega)|_{\omega\to 0}$, on the driving peak electric field. These data are taken for narrowband apical oxygen driving at 19.2 THz, at T = 100 K. The linear electric field dependence, without any signature of saturation, confirms and extends the results reported in Ref. 5.

All data reported in Figs. 2,3,4,5 of the main text have been taken at the maximum electric field of ~3 MV/cm, corresponding to a pump fluence of ~8 mJ/cm².

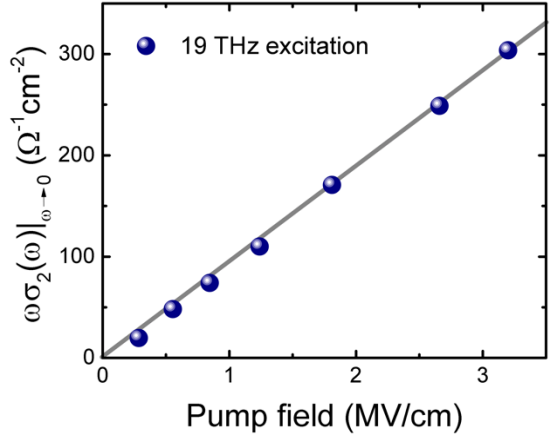

**Figure S7**. Pump electric field dependence of the coherent component of the light-induced response, represented by the low-frequency limit $\omega\sigma_2(\omega)|_{\omega\to 0}$, measured in YB$_2$Cu$_3$O$_{6.5}$ at 100 K after photo-excitation with narrowband pulses at 19.2 THz.

# S8. Transient response below T$_C$

The transient response in the superconducting state of YBa$_2$Cu$_3$O$_{6.5}$ was investigated at $T = 10$ K for the same excitation frequencies, 4.2 THz and 19.2 THz, for which we have reported the temperature dependent response in Fig. 5 of the main text.

In Figure S8 we report the dynamical evolution of the real and imaginary part of the complex optical conductivity for 4.2 THz driving at ~8 mJ/cm² fluence, 19.2 THz driving at ~8 mJ/cm² fluence, and 19.2 THz driving at ~1.5 mJ/cm² fluence.

For apical oxygen excitation at 19.2 THz with high fluence (middle panel) we observe an analogous response to that reported previously [4,5], namely a short-lived enhancement in the $\sigma_2(\omega)$ divergence after photo-excitation, accompanied by a weak offset in $\sigma_1(\omega)$. This behavior is indicative of a transient increase in superfluid density, which (in analogy with the observation above T$_C$) appears to be stronger than in previous experiments due to the higher available pump fluence.

If the same experiment is repeated at a lower pump fluence (right panel), comparable to that employed in Ref. 11, only a transient depletion of the superconducting condensate is observed.

Finally, for 4.2 THz pump (left panel) a similar depletion and no enhancement is found for any driving field, a result which shall be related to the absence of any light-induced coherence above $T_C$ at this excitation frequency (see Figs. 3 & 4 of main text).

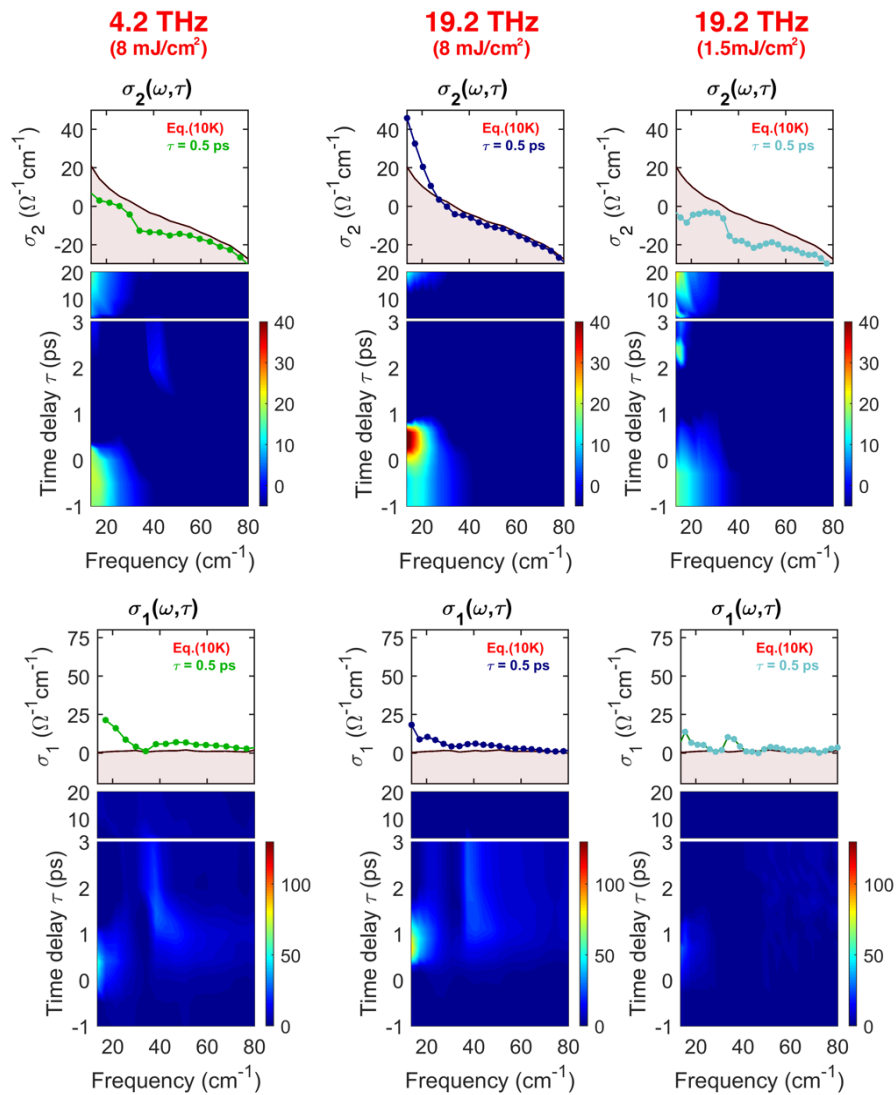

**Figure S8**. Frequency- and time-delay-dependent real and imaginary part of the optical conductivity measured in YBa$_2$Cu$_3$O$_{6.5}$ at T = 10 K for narrowband excitation at 4.2 THz (left panel) and 19.2 THz (at two pump fluences, center and right panels). Upper panels: corresponding line cuts displayed at equilibrium (grey lines) and at the time delay corresponding to the peak of the response (colored circles).

## S9. Pump-induced heating

Following the procedure described in Ref. 12, we have used a two-temperature model to estimate the lattice temperature after electron-phonon thermalization, induced by photoexcitation at two different pump frequencies.

The specific heat for YBa$_2$Cu$_3$O$_{6.5}$ is described by the relation $C_S = \gamma T + \beta T^3$ over a wide temperature range, where $\gamma \simeq 2.3$ mJ/mol$^{-1}$K$^{-2}$ and $\beta \simeq 0.394$ mJ/mol$^{-1}$K$^{-4}$ are the electronic and lattice coefficients to the specific heat estimated from literature [13], respectively.

For a given pump fluence, we calculated the absorbed energy over the photo-excited region and determined the effective temperature after electron phonon-thermalization by integrating $Q_{pump} = \int_{T_i}^{T_f} NC_S(T)dT$, where $Q_{pump}$ is the total energy absorbed from the pump pulse, $N$ is the number of moles in the excited volume, $T_i$ is the initial temperature and $T_f$ the temperature after electron-phonon thermalization. The absorbed energy was estimated by $Q_{pump} = FA(1-R)$, where $F$ is the pump fluence, $A$ is the area of the pump beam on the sample, and $R$ is the reflectivity at the pump frequency. The excited volume was calculated as a cylindrical disk with a diameter of ~400 μm and height equal to the pump penetration depth.

The final temperature, $T_f$, is plotted versus pump fluence in Fig. S9 for an initial temperature $T_i$ = 100 K. The colored circles correspond to the highest fluence used in our experiment, for which the pump-induced heating is limited to below ~20 K.

We stress here that our paper focuses on the analysis of early time delay dynamics ($\tau \lesssim 3$ ps), for which full thermalization has certainly not occurred yet [14]. The calculation reported here is relevant for dynamics at longer time delays, which is not the main subject of the present work.

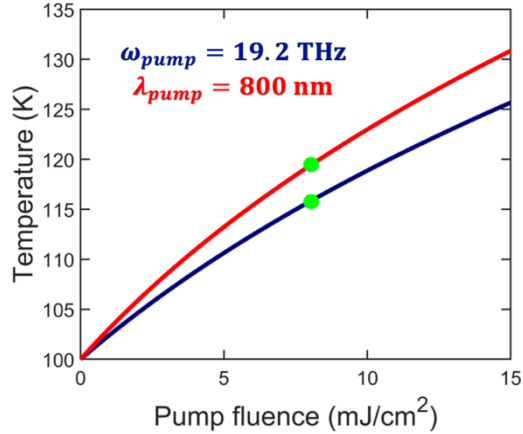

**Figure S9**. Calculated final lattice temperature after electron-phonon thermalization as a function of excitation fluence for 19 THz (blue) and 800 nm pump (red). The same approach reported in Ref. 12 has been applied. The maximum pump fluence used in our experiment is indicated by green dots.

## S10. Ab-initio calculations of the structural dynamics

In this Section, we correlate the experimentally-determined frequency band for the stimulation of transient superconductivity with the optically-driven rearrangement of the crystal structure, which we calculated by combining effective Hamiltonian modeling with first-principle computations.

Our approach is based on an anharmonic crystal potential that consists of three distinct contributions [15,16,17]:

1. The harmonic potential of each phonon mode

$$V_{harm} = \sum \frac{\omega_i^2}{2} Q_i^2, \quad (S10.1)$$

with $\omega_i$ and $Q_i$ representing the eigenfrequency and coordinate of the $i$-th mode, respectively.

2. The anharmonic potential containing higher-order terms of the phonon coordinates and combinations of different phonon modes

$$V_{anharm} = \sum g_{ijk} Q_i Q_j Q_k + \sum f_{iklm} Q_i Q_k Q_l Q_m, \quad (S10.2)$$

with $g_{ijk}$ and $f_{iklm}$ indicating third and fourth order anharmonic coefficients, respectively.

3. The coupling of each individual phonon mode to an external electric field

$$V_{efield} = \sum Z_i^* Q_i E_{field}, \quad (S10.3)$$

with $Z_i^*$ representing the mode effective charge [18,19].

The structural dynamics are then determined by the equations of motion for each phonon mode, given by

$$\ddot{Q}_i + 2\gamma_i \dot{Q}_i + \nabla_{Q_i}(V_{harm} + V_{anharm} + V_{efield}) = 0. \quad (S10.4)$$

Here, we introduced a phenomenological damping term $\gamma_i$, which accounts for contributions to the finite lifetime which are not already considered within the anharmonic potential. Importantly, the equations are restricted to phonon modes at the Brillouin zone center, due to the long wavelengths of the THz excitation pulses.

Various studies have shown that this approach is able to describe the structural dynamics induced by the resonant excitation of infrared-active modes in a solid [20,21]. However, only a small subset of all possible phonon coupling constants was considered in these cases, predominantly due to computational limitations. In the present simulations, this reduction was overcome by utilizing the approaches given in Refs. [15,22]. We list all technical and numerical details at the end of this Section.

We simulated the THz-driven structural dynamics of the ortho-II structure of $YBa_2Cu_3O_{6.5}$, which exhibits 73 non-translational phonon modes at the Brillouin zone center. The most relevant phonon modes for $c$-axis polarized THz excitation are 13 $B_{1u}$ modes, which exhibit a finite electric dipolar moment along this direction. In addition, there are 11 $A_g$ modes, which fulfill the symmetry requirements to exhibit a finite third-order type coupling to the $B_{1u}$ modes [15,20]. A full list of the eigenfrequencies of these modes is presented at the end of this Section.

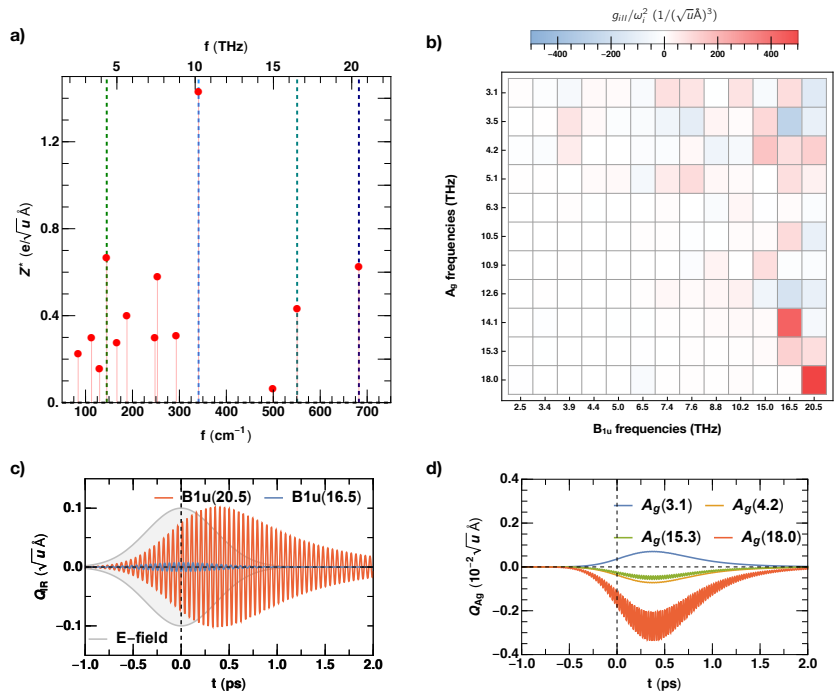

**Figure S10.1**. **a.** Mode effective charges for the polar $B_{1u}$ modes of the $YBa_2Cu_3O_{6.5}$ ortho-II structure. **b.** Third-order anharmonic coupling coefficients of the type $Q_{Ag}Q_{IR}^2$ for all $B_{1u}$ and $A_g$ modes. **c.** Oscillations of two $B_{1u}$ modes at 16.5 and 20.5 THz driven by a narrowband THz pulse at 20.5 THz (grey area). **d.** Dynamics of the four $A_g$ modes exhibiting stronger coupling to the 20.5-THz $B_{1u}$ mode driven in panel c.

The calculated effective charges for the $B_{1u}$ modes, needed to describe their excitation by a THz electric field according to Eq. (S10.3), are shown in Figure S10.1a. We assumed here that the Born effective charge of each atom equals its ionic charge. In agreement with experimental observations [23], we obtain large values for the two highest-frequency apical-oxygen modes at 16.5 THz and 20.5 THz, for the 10-THz mode affecting the $CuO_2$ layers, and for the mode at 4 THz that involves Ba displacements. Note also that all phonon frequencies match within a few percent the values determined from the experiment.

Group symmetry dictates three sets of third-order couplings for the phonon modes considered. The first is the coupling of the square of the optically-excited $B_{1u}$ modes to a single $A_{1g}$ mode, proportional to $Q_{IR,i}^2 Q_{Ag}$. The second is the coupling of two different $B_{1u}$ modes to a single $A_{1g}$ mode ($Q_{IR,i} Q_{IR,j} Q_{Ag}$), and the third is the coupling between three $A_{1g}$ modes ($Q_{Ag,i} Q_{Ag,j} Q_{Ag,k}$). For the fourth-order coupling terms, we only considered the terms of the form $Q_i^4$ and neglected expressions mixing two or more phonon modes. Thus, we took into account, in total, 1067 third-order and 24 fourth-order coupling coefficients. Note that, in the earlier studies of Refs. [15,20], at maximum 24 coupling coefficients were considered. As an example, Figure S10.1b shows a subset of the computed anharmonic coupling coefficients, representing the two-mode coupling of the type $Q_{IR,i}^2 Q_{Ag}$. Both the two highest-frequency $B_{1u}$ polar modes couple strongly to two different $A_{1g}$ modes, similar to the observations reported for $YBa_2Cu_3O_7$ in Ref. [15].

Figure S10.1c,d show the dynamics of two selected $B_{1u}$ modes and four anharmonically coupled $A_g$ modes. The time traces have been triggered by an electric field pulse $E(t) = e^{\frac{2t^2 ln(2)}{FWHM^2}} cos(\omega t)$ with a 3 MV/cm peak field and 0.6 ps pulse duration (FWHM) at 20.5 THz. The resonantly excited 20.5-THz $B_{1u}$ mode oscillates with an amplitude that is about one order of magnitude larger than that of the off-resonant $B_{1u}$ mode at 16.5 THz. The dominant $A_g$ modes, anharmonically coupled to the $B_{1u}$ phonons, are transiently displaced from their equilibrium positions with significant amplitudes ($|Q_{Ag}| < 10^{-5}\sqrt{u}$Å). They reach their maximum displacements at about 0.4 ps time delay, for which also the resonantly driven 20.5-THz $B_{1u}$ mode reaches its peak amplitude.

In the same way, we constructed the time dependent crystal structures for different excitation frequencies between 2.5 and 22 THz and focused on the transient displacements of $A_g$ modes, which involve an average distortion of the crystal lattice [15,20]. Figure S10.2 shows the maximum changes of selected bonds (distance of the apical oxygen atoms from the $CuO_2$ planes, inter-bilayer distance of the $CuO_2$ planes, and buckling of the $CuO_2$ planes) calculated from these average phonon displacements. Peak electric field and pulse duration were kept fixed at 3 MV/cm and 0.6 ps, respectively, for all driving frequencies.

For excitation in the interval between 5 and 10 THz, we observed a significant shift of the apical oxygen atoms closer to the $CuO_2$ planes, accompanied by a buckling within the $CuO_2$ layers and an increase in the intra-bilayer distance. In contrast, very different rearrangements were found for driving frequencies within the range for which photo-induced superconducting coherence was observed. Most prominently, the apical oxygen atoms moved away from the $CuO_2$ planes for resonant driving of the 20.5-THz phonon, while excitation of the 16.5-THz mode shifted the atoms closer to the planes. In both cases, we also calculated different signs for the $CuO_2$ plane buckling and the intra-bilayer distance, which increased for 16.5-THz but decreased for the 20.5-THz excitation.

Overall, we did not find a consistent trend in the transient crystal structures, which was unique to the frequency band for which transient superconductivity was observed. As such, this result puts some doubts on the significance of anharmonic phonon coupling as a possible mechanism for optically-driven superconductivity in YBa$_2$Cu$_3$O$_{6+x}$, as posited earlier in Ref. [20].

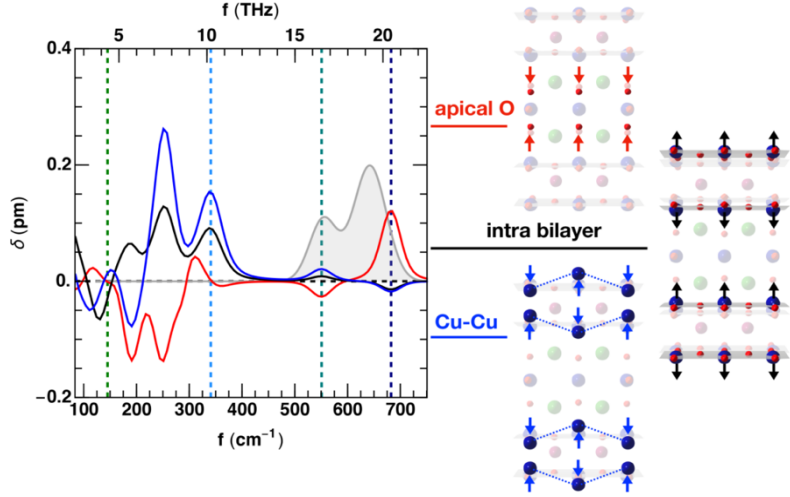

**Figure S10.2.** Transient average distortions of the apical oxygen positions (red), the intra-bilayer distance (black) and the planar Cu buckling (blue) in YBa$_2$Cu$_3$O$_{6.5}$, derived from ab-initio calculations of nonlinear phonon-phonon interactions for different excitation frequencies. The grey shaded area depicts the experimentally determined frequency range of the optically-enhanced superconducting response shown in Figure 4 of the main text. The excitation parameters of 600 fs pulse duration and 3 MV/cm peak electric field, used in the calculations, reproduce the experimental conditions. Arrows in the illustrations shown on the right correspond to positive displacements.

***Technical and numerical approach for the ab-initio calculations:***

We employed first-principles total energy calculations in the framework of the density functional theory (DFT) to compute all harmonic and anharmonic terms included in the equations (S10.1), (S10.2), and (S10.3). Specifically, we used the implantation of DFT, applying the linearized augmented-plane wave method (LAPW) within the ELK-code [24]. We approximated the exchange-correlation functional by the local density approximation. In addition, we performed careful tests of all relevant numerical parameters entering the computation. The setting for well-converged results corresponded to a truncation at $l_{max}$=10 of the angular expansion of wave functions and potential within the muffin-tin radii (2.6, 2.8, 1.85 and 1.4 a.u. for Y, Ba, Cu, and O, respectively). A $|G|_{max}$=20 a.u.$^{-1}$ limited the potential and density expansion within the interstitial region. We set $R_{MT} \times k_{max}$=8.0 for truncating the plane-wave wavefunction expansion. The Brillouin zone was sampled within our computations by a 11×19×5 k-point mesh. The same configuration was also employed in a previous study [25] and, within this setting, the forces (most relevant for phonons and anharmonic terms) were made to converge by less than 0.1 meV/Å.

Before computing the coefficients of the anharmonic potential, we first structurally relaxed the unit cell of $YBa_2Cu_3O_{6.5}$, for which we considered the ortho-II structure, following previous studies and the experimental setting [20]. As lowest energy state, we obtained the atomic configuration given in Table S10.1.

Finally, the phonon eigensystem was computed by frozen-phonon calculation using symmetry-adapted distortions generated with the Phonopy package [26]. We list all modes at the zone center in Table S10.2. Next, we applied the prescription in Ref. [22] to calculate the anharmonic constants of third order and took the approach of Ref. [15] for the quartic order terms.

Last, we computed the mode-effective charges appearing in equation (S10.3) by using the nominal averaged ionic charges of each atom as Born charges. Explicitly, we have for Yttrium 3+, Barium 2+, Copper 2+ and Oxygen 2-.

**Table S10.1.** DFT minimized structural configuration of the $YBa_2Cu_3O_{6.5}$ ortho-II cell with $a = 7.55$ Å, $b = 3.81$ Å, and $c = 11.50$ Å.

| element | Wykoff position | x | z | element | Wykoff position | x | z |
|---|---|---|---|---|---|---|---|
| Y | l | 0.251 | 0.500 | O1 | e | 0.000 | 0.000 |
| Ba | x | 0.244 | 0.187 | O2 | w | 0.250 | 0.378 |
| Cu1 | a | 0.000 | 0.000 | O3 | r | 0.000 | 0.378 |
| Cu2 | b | 0.500 | 0.000 | O4 | t | 0.500 | 0.378 |
| Cu3 | q | 0.000 | 0.356 | O5 | q | 0.000 | 0.161 |
| Cu4 | s | 0.500 | 0.355 | O6 | s | 0.500 | 0.153 |

**Table S10.2.** Computed DFT eigenfrequencies of the phonon modes at the center of the Brillouin zone for the $YBa_2Cu_3O_{6.5}$ ortho-II structure.

| Label | f (THz) | Label | f (THz) | Label | f (THz) | Label | f (THz) |
|---|---|---|---|---|---|---|---|
| $A_g$ | 3.1 | $B_{1u}$ | 7.4 | $B_{2u}$ | 17.0 | $B_{3u}$ | 8.1 |
| $A_g$ | 3.5 | $B_{1u}$ | 7.6 | $B_{2u}$ | 17.4 | $B_{3u}$ | 8.4 |
| $A_g$ | 4.2 | $B_{1u}$ | 8.8 | $B_{2g}$ | 1.8 | $B_{3u}$ | 10.3 |
| $A_g$ | 5.1 | $B_{1u}$ | 10.2 | $B_{2g}$ | 2.6 | $B_{3u}$ | 10.5 |
| $A_g$ | 6.3 | $B_{1u}$ | 15.0 | $B_{2g}$ | 4.1 | $B_{3u}$ | 11.8 |
| $A_g$ | 10.5 | $B_{1u}$ | 16.5 | $B_{2g}$ | 4.2 | $B_{3u}$ | 13.4 |
| $A_g$ | 10.9 | $B_{1u}$ | 20.5 | $B_{2g}$ | 6.5 | $B_{3u}$ | 18.3 |
| $A_g$ | 12.6 | $B_{1g}$ | 2.9 | $B_{2g}$ | 7.4 | $B_{3g}$ | 2.0 |
| $A_g$ | 14.1 | $B_{1g}$ | 3.8 | $B_{2g}$ | 8.7 | $B_{3g}$ | 4.1 |
| $A_g$ | 15.3 | $B_{1g}$ | 9.7 | $B_{2g}$ | 10.0 | $B_{3g}$ | 5.4 |
| $A_g$ | 18.0 | $B_{2u}$ | 2.4 | $B_{2g}$ | 11.3 | $B_{3g}$ | 6.5 |
| $A_u$ | 2.8 | $B_{2u}$ | 3.6 | $B_{2g}$ | 11.4 | $B_{3g}$ | 10.3 |
| $A_u$ | 10.8 | $B_{2u}$ | 4.7 | $B_{2g}$ | 18.1 | $B_{3g}$ | 11.1 |
| $B_{1u}$ | 2.5 | $B_{2u}$ | 5.2 | $B_{3u}$ | 2.5 | $B_{3g}$ | 16.6 |
| $B_{1u}$ | 3.4 | $B_{2u}$ | 5.6 | $B_{3u}$ | 2.9 | $B_{3g}$ | 17.2 |
| $B_{1u}$ | 3.9 | $B_{2u}$ | 8.0 | $B_{3u}$ | 3.7 | | |
| $B_{1u}$ | 4.4 | $B_{2u}$ | 10.3 | $B_{3u}$ | 3.8 | | |
| $B_{1u}$ | 5.0 | $B_{2u}$ | 11.5 | $B_{3u}$ | 4.7 | | |
| $B_{1u}$ | 5.6 | $B_{2u}$ | 16.6 | $B_{3u}$ | 5.4 | | |

# References (Supplemental Material)

[1] A. Sell, A. Leitenstorfer, and R. Huber, Opt. Lett. **33**, 2767 (2008).

[2] C. Manzoni, H. Ehrke, M. Först, and A. Cavalleri, Opt. Lett. **35**, 757 (2010).

[3] B. Liu, H. Bromberger, A. Cartella, T. Gebert, M. Först, and A. Cavalleri, Opt. Lett. **42**, 129 (2017).

[4] W. Hu, S. Kaiser, D. Nicoletti, C. R. Hunt, I. Gierz, M. C. Hoffmann, M. Le Tacon, T. Loew, B. Keimer, and A. Cavalleri, Nat. Mater. **13**, 705 (2014).

[5] S. Kaiser, C. R. Hunt, D. Nicoletti, W. Hu, I. Gierz, H. Y. Liu, M. Le Tacon, T. Loew, D. Haug, B. Keimer, and A. Cavalleri, Phys. Rev. B **89**, 184516 (2014).

[6] D. Nicoletti, D. Fu, O. Mehio, S. Moore, A. S. Disa, G. D. Gu, and A. Cavalleri, Phys. Rev. Lett. **121**, 267003 (2018).

[7] C. C. Homes, T. Timusk, D. A. Bonn, R. Liang, and W. N. Hardy, Physica C **254**, 265-280 (1995).

[8] E. Uykur, K. Tanaka, T. Masui, S. Miyasaka, and S. Tajima, Journal of Physical Society of Japan **81**, SB035 (2012).

[9] E. Uykur, "Pseudogap and precursor superconductivity study of $YBa_2(Cu_{1-x}Zn_x)_3O_y$: c-axis optical study from underdoped to overdoped region", Springer Japan (2015).

[10] D. van der Marel and A. Tsvetkov, Czech. J. Phys. **46**, 3165 (1996).

[11] S. J. Zhang, Z. X. Wang, H. Xiang, X. Yao, Q. M. Liu, L. Y. Shi, T. Lin, T. Dong, D. Wu, N. L. Wang, *arXiv:1904.10381* (2019).

[12] K. A. Cremin, J. Zhang, C. C. Homes, G. D. Gu, Z. Sun, M. M. Fogler, A. J. Millis, D. N. Basov, R. D. Averitt, *Proc. Natl. Acad. Sci. U.S.A.* **116**, 19875-19879 (2019).

[13] J. B. Kemper, Measurement of the heat capacity of cuprate superconductors in high magnetic fields. *Doctoral Thesis*, College of arts and science, Florida State University (2014).

[14] H. Niwa, N. Yoshikawa, K. Tomari, R. Matsunaga, D. Song, H. Eisaki, and R. Shimano, Phys. Rev. B **100**, 104507 (2019).

[15] M. Fechner and N. A. Spaldin, Phys. Rev. B **94**, 134307 (2016).

[16] A. Subedi, A. Cavalleri, and A. Georges, Phys. Rev. B **89**, 220301 (2014).

[17] D. M. Juraschek, M. Fechner, and N. A. Spaldin, Phys. Rev. Lett. **118**, 054101 (2017).

[18] X. Gonze and C. Lee, Phys. Rev. B **55**, 10355 (1997).

[19] A. Cartella, T. F. Nova, M. Fechner, R. Merlin, and A. Cavalleri, Proc. Natl. Acad. Sci. USA **115**, 12148 (2018).

[20]    R. Mankowsky, A. Subedi, M. Först, S. O. Mariager, M. Chollet, H. T. Lemke, J. S. Robinson, J. M. Glownia, M. P. Minitti, A. Frano, M. Fechner, N. A. Spaldin, T. Loew, B. Keimer, A. Georges, and A. Cavalleri, Nature **516**, 71 (2014).

[21]    M. Kozina, M. Fechner, P. Marsik, T. van Driel, J. M. Glownia, C. Bernhard, M. Radovic, D. Zhu, S. Bonetti, U. Staub, and M. C. Hoffmann, Nat. Phys. **15**, 387 (2019).

[22]    G. Khalsa and N. A. Benedek, Npj Quantum Mater. **3**, 15 (2018).

[23]    C. C. Homes, T. Timusk, D. A. Bonn, R. Liang, and W. N. Hardy, Can. J. Phys. **675**, 663 (1995).

[24]    http://elk.sourceforge.net, (n.d.).

[25]    R. Mankowsky, M. Fechner, M. Först, A. von Hoegen, J. Porras, T. Loew, G. L. Dakovski, M. Seaberg, S. Möller, G. Coslovich, B. Keimer, S. S. Dhesi, and A. Cavalleri, Struct. Dyn. **4**, 044007 (2017).

[26]    A. Togo and I. Tanaka, Scr. Mater. **108**, 1 (2015).